\documentclass[superscriptaddress,prd,aps,showpacs,nofootinbib,showkeywords,eqsecnum,preprint]{revtex4-1}
\usepackage{graphicx,color,amsmath,amsxtra}

\usepackage[final]{hyperref} 
\hypersetup{
	colorlinks=true,       
	linkcolor=red,        
	citecolor=green,        
	filecolor=magenta,     
	urlcolor=magenta     
}
\usepackage{epsf}
\usepackage{amssymb}
\usepackage{enumerate}
\usepackage{hhline}
\usepackage{array}
\usepackage{tabularx}
\usepackage{graphicx}                
\usepackage{epstopdf}
\usepackage{caption}
\usepackage{subcaption}
 \usepackage{multirow}
 \usepackage{booktabs}

\newcommand{\Lagr}{\mathcal{L}}

\begin{document}
\pagestyle{myheadings}
\title{Stability investigations of  de Sitter inflationary solutions in power-law extensions of the Starobinsky model}
\author{Tuyen M. Pham}
\email{tuyen.phammanh@phenikaa-uni.edu.vn}
\affiliation{Phenikaa Institute for Advanced Study, Phenikaa University, Hanoi 12116, Vietnam}
\author{Duy H. Nguyen}
\email{duy.nguyenhoang@phenikaa-uni.edu.vn}
\affiliation{Phenikaa Institute for Advanced Study, Phenikaa University, Hanoi 12116, Vietnam}
\author{Tuan Q. Do}
\email{tuan.doquoc@phenikaa-uni.edu.vn}
\affiliation{Phenikaa Institute for Advanced Study, Phenikaa University, Hanoi 12116, Vietnam}
\affiliation{Faculty of Basic Sciences, Phenikaa University, Hanoi 12116, Vietnam}
\author{W. F. Kao}
\email{homegore09@nycu.edu.tw}
\affiliation{Institute of Physics, National Yang Ming Chiao Tung University, Hsin Chu 30010, Taiwan}
\date{\today} 

\begin{abstract}
In this paper, we would like to examine whether stable de Sitter inflationary solutions appear within power-law extensions of the Starobinsky model. In particular, we will address general constraints for the existence along with the stability of de Sitter inflationary solutions in a general case involving not only the Starobinsky $R^2$ term but also an additional power-law $R^n$ one. According to the obtained results, we will be able to identify which extension is more suitable for an early inflationary phase rather than a late-time cosmic acceleration phase. To be more specific, we will consider several values of $n$ to see whether the corresponding de Sitter inflationary solutions are stable or not.
\end{abstract}
\maketitle
\section{Introduction} \label{intro} 
Cosmic inflation has played a leading paradigm in modern cosmology \cite{Starobinsky:1980te,Guth:1980zm,Linde:1981mu,Linde:1983gd}. It turns out that theoretical predictions based on cosmic inflation have been well consistent with the observational data of the leading cosmic microwave background (CMB) radiation probes such as the WMAP \cite{Hinshaw:2012aka} or Planck satellite \cite{Aghanim:2018eyx,Akrami:2018odb}. Remarkably, although an inflationary phase is widely believed to be caused by the so-called inflaton field, which is nothing but a scalar field, the Starobinsky model, which is one of the first inflationary models and does not have such inflaton field, has still been one of the most viable inflationary models in the light of the latest data of the Planck \cite{Aghanim:2018eyx,Akrami:2018odb}. 

In principle, the Starobinsky model is a non-trivial extension of the Einstein gravity, in which a quadratic correction term $R^2$ is added into the Einstein-Hilbert action as shown in Ref. \cite{Starobinsky:1980te}. Starobinsky pointed out in this seminal paper that this $R^2$ term is nothing but a quantum correction of Einstein gravity. By using a suitable conformal transformation, the Starobinsky model can be written as an effective model of inflaton \cite{Whitt:1984pd,Maeda:1987xf,Barrow:1988xh} and therefore its CMB imprints can be worked out accordingly for comparisons with observations, e.g., see Refs. \cite{Mishra:2018dtg,Mishra:2019ymr} for detailed calculations. See also Ref. \cite{Maeda:1988ab} for conformal transformations of extensions of Starobinsky model, e.g., a model of arbitrary polynomial of $R$.

Among the higher-order gravity models, the Starobinsky model is a unique one due to not only its great success  in dealing with the inflationary phase but also its freedom from the so-called Ostrogradsky instability \cite{Woodard:2015zca}. Moreover, the Starobinsky model is among higher-order curvature gravity models, which have been regarded as promising candidates for seeking an ultraviolet (UV) completeness of Einstein's general relativity \cite{Koshelev:2017tvv}. This fact is based on a seminal work of Stelle in Ref. \cite{Stelle:1976gc}, in which he pointed out that a renormalizable gravity model could be archived once the quadratic curvature correction terms such as $R^2$ and $R_{\mu\nu}R^{\mu\nu}$ are introduced into the pure Einstein-Hilbert action. And very interestingly, the Starobinsky model belongs to a class of fourth-order gravity ones, whose field equations involve fourth-order derivatives coming from quadratic curvature terms \cite{Starobinsky:1987zz,Muller:1987hp,Muller:1989rp}. In addition to the inflationary phase, fourth-order gravities would have provided us reasonable resolutions to explain the accelerated expansion phase of our universe \cite{Carroll:2004de,Schmidt:2006jt,Ivanov:2011vy,DeLaurentis:2015fea,Salvio:2018crh}. Last but not least, the Starobinsky model is a typical model of $f(R)$ gravity theory, whose rich cosmological implications can be found in interesting review papers, e.g., see Refs. \cite{Sotiriou:2008rp,DeFelice:2010aj,Capozziello:2011et,Nojiri:2010wj,Nojiri:2017ncd,Odintsov:2023weg}. 
 
Very soon after the appearance of the Starobinsky model, many non-trivial extensions of this model have been proposed in various directions, see Refs. \cite{Barrow:1983rx,Madsen:1989rz,Berkin:1990nu,Mignemi:1991wa,Wands:1993uu,Kluske:1995vj,Clifton:2005at,Barrow:2005qv,Clifton:2006kc,Barrow:2006xb,Toporensky:2006kc,Barrow:2009gx,Appleby:2009uf,Middleton:2010bv,Iihoshi:2010pf,Myrzakulov:2014hca,Netto:2015cba,Myrzakulov:2016tsz,Elizalde:2017mrn,Muller:2017nxg,Liu:2018hno,Aldabergenov:2018qhs,Elizalde:2018now,Elizalde:2018rmz,Koshelev:2022olc,Ketov:2022lhx,Ketov:2022zhp,Pozdeeva:2022lcj,Ketov:2010qz,Ketov:2012se,Farakos:2013cqa,Ferrara:2013kca,Ferrara:2013wka,Watanabe:2013lwa,Huang:2013hsb,Sebastiani:2013eqa,Cheong:2020rao,Cano:2020oaa,Rodrigues-da-Silva:2021jab,Ivanov:2021chn,Shtanov:2022pdx,Modak:2022gol,Do:2023yvg} for an incomplete list of literature. Among these extensions, we are now interested in non-trivial extensions of the Starobinsky model (e.g., see Refs. \cite{Ketov:2010qz,Ketov:2012se,Farakos:2013cqa,Ferrara:2013kca,Ferrara:2013wka,Watanabe:2013lwa,Huang:2013hsb,Sebastiani:2013eqa,Cheong:2020rao,Cano:2020oaa,Rodrigues-da-Silva:2021jab,Ivanov:2021chn,Shtanov:2022pdx,Modak:2022gol} for recent related investigations and Refs. \cite{Barrow:1988xh,Barrow:1983rx,Madsen:1989rz} for earlier related discussions), in which an additional higher-order term $R^n$ ($n\neq 0,~1$, and $2$) is added along with the well-known  term $R^2$ into the pure Einstein-Hilbert action. It has been shown in Refs. \cite{Ketov:2010qz,Ketov:2012se,Farakos:2013cqa,Ferrara:2013kca,Ferrara:2013wka} that  a higher-order term $R^n$ may be originated from the supergravity.  These extensions will be dubbed power-law extensions due to the existence of power-law function of Ricci scalar, i.e., $R^n$, in this paper for convenience. More specifically, we will investigate on the existence and stability of de Sitter inflationary solutions in these extensions, following our previous work in Ref. \cite{Do:2023yvg} as well as many previous studies on the similar issue, e.g., see Refs. \cite{Toporensky:2006kc,Appleby:2009uf,Netto:2015cba,Ketov:2022zhp,Ivanov:2011np,Skugoreva:2014gka,Elizalde:2014xva,Pozdeeva:2019agu,Vernov:2021hxo}. It turns out that the stability of de Sitter inflationary solutions found in power-law extensions of the Starobinsky model  is really an important information because it would tell us which phase, early time or late time, of our universe is more compatible with these extensions. In particular, if a found de Sitter inflationary solution turns out to be unstable then the corresponding power-law extension of the Starobinsky model would be suitable for the inflationary phase of the universe \cite{Pozdeeva:2019agu,Vernov:2021hxo}. Similar conclusion would also be valid for a case of non-existence of de Sitter inflationary solution \cite{Pozdeeva:2019agu,Vernov:2021hxo}. It is important to note that the original Starobinsky model \cite{Starobinsky:1980te} does not admit any exact de Sitter inflationary solution as confirmed in many previous works, e.g., see Refs. \cite{Ketov:2022zhp,Do:2023yvg}. Instead, it was shown to admit a quasi-de Sitter inflationary solution \cite{Starobinsky:1980te}. This result could be a useful criteria for checking whether non-trivial extensions of Starobinsky model are relevant to the inflationary phase. For example, it has been shown in Refs. \cite{Ketov:2022zhp,Do:2023yvg} that a de Sitter inflationary solution of the so-called Starobinsky-Bel-Robinson gravity is indeed unstable and therefore this extension would have a chance to be a viable inflationary model \cite{Ketov:2022zhp}. On the other hand, if a found de Sitter inflationary solution turns out stable then the corresponding power-law extension of the Starobinsky model would be suitable for the late time phase of the universe, which has been undergone an accelerated expansion \cite{Pozdeeva:2019agu,Vernov:2021hxo}. This is due to an issue that a stable de Sitter inflationary solution could be a basis of the so-called eternal inflation \cite{Guth:2007ng}, which must require a graceful exit to ensure a smooth transition between the inflationary phase and a late time expanding phase of our universe. Unfortunately, figuring out a suitable graceful exit for this eternal inflation is not always straightforward \cite{Brustein:1994kw}. 

It should be mentioned that de Sitter solutions were investigated in a polynomial gravity with a general Lagrangian $L_g =f(R)=\sum_{n=0}^N a_n R^n$ ($N \in \mathbb{Z}^+$) by Barrow and his colleagues quite a long time ago \cite{Barrow:1988xh,Barrow:1983rx,Madsen:1989rz}. Very recently, many people have investigated several power-law extensions of the Starobinsky model of inflation \cite{Cheong:2020rao,Cano:2020oaa,Rodrigues-da-Silva:2021jab,Ivanov:2021chn,Shtanov:2022pdx,Modak:2022gol}. In harmony with these interesting works, therefore, we would like to examine whether (un)stable de Sitter inflationary solutions exist in power-law extensions of the Starobinsky model. More specifically, we will derive general constraints for the existence of de Sitter inflationary solution within a general power-law extension of the Starobinsky model with the corresponding Lagrangian given by ${\cal L} =\sqrt{-g}\left[R+\beta_1 R^2 +\beta_2 R^n \right]$ with $n\in \mathbb{R}$ and $n \neq 0, 1, 2$. Then, we will investigate the stability of the obtained de Sitter solution for several $n$ during the inflationary phase using the well-known dynamical system method. Due to the stability of these solutions, we might be able to finalize which phase of our universe is the most relevant to them. 

 In summary, this paper will be organized as follows: (i) Its brief introduction has been written in Sec. \ref{intro}. (ii) Basic setup of power-law extension of the Starobinsky model will be presented in Sec. \ref{sec2}. (iii) The corresponding de Sitter inflationary solutions will be solved in Sec. \ref{sec3}. (iv) The corresponding dynamical system along with its fixed point will be written in Sec. \ref{sec4}. (v) Stability of the obtained inflationary solutions will be analyzed in Sec. \ref{sec5}. (vi) Some specific models will be examined in Sec. \ref{sec6}.  {\color{black}(vii) Additional stability analysis using an effective potential method, which acts as a cross-check of the stability analysis in Sec. \ref{sec6}, will be presented in Sec. \ref{sec7}.} Finally, concluding remarks will be given in Sec. \ref{final}. 
\section{Basic setup and field equations} \label{sec2}
In order to maintain the role of the $R^2$ term, we will consider the following extension with a general action given by
	\begin{equation}\label{1}
		S =\frac{M_p^2}{2} \int d^4 x  \sqrt{-g}  \left[R+\beta_1 R^2+ \beta_2 \hat f (R)\right],
	\end{equation}
	where $\beta_1$ and $\beta_2$ are all non-vanishing parameters. It is noted that the coefficient of $R$ term has been fixed to be one as usual, i.e., $\beta_0 =1$. In addition, $M_p$ is the reduced Planck mass, while $\hat f (R)$ is an undetermined function of the Ricci scalar $R$. It is clear that once $\beta_2 \to 0$ then the above action will reduce to the well-known action of original Starobinsky model, in which $\beta_1 \equiv 1/(6m^2)>0$ with $m$ being the mass of inflaton (a.k.a. the scalaron mass) \cite{Starobinsky:1980te,Ivanov:2021chn}. However, we would like to let $\beta_1$ be a free parameter in this paper for a panorama picture, following our recent work \cite{Do:2023yvg}. Indeed, it has been shown in Ref. \cite{Do:2023yvg} that we still obtain de Sitter inflationary solutions for negative $\beta_1$ in the presence of additional correction terms such as the Bel-Robinson one.
	
	Ones could of course claim that the above action is nothing but that of $f(R)$ gravity model. However, it should be careful that not all $f(R)$ models contain the $R^2$ term. It appears that some power-law types of $\hat f (R)$ have been investigated in a recent paper \cite{Ivanov:2021chn} such as $\hat f(R)=R^{3/2}, R^3$, and $R^4$.  It is worth noting that the authors of Ref.  \cite{Ivanov:2021chn} have arrived an important conclusion that unlike the $R^3$ and $R^4$  terms, the term $R^{3/2}$ has a significant impact on the value of tensor-to-scalar ratio.

In this paper, we will focus on the homogeneous and isotropic Friedmann-Lemaitre-Robertson-Walker  (FLRW) spacetime, whose element is given by
\begin{equation}\label{2}
ds^2 = -N(t)^2 dt^2+e^{2\alpha(t)}\left(dx^2+dy^2+dz^2 \right),
\end{equation}
where $N(t)$ is the lapse function introduced to derive the corresponding Hamiltonian constraint \cite{Myrzakulov:2014hca,Toporensky:2006kc,Do:2023yvg,Do:2020vdc,Do:2021fal,Kao:1991zz}. And it should be set to be one after field equations are derived. In addition, $\alpha(t)$ is a scale factor function depending on a cosmic time $t$. For expanding universes, $\alpha(t)$ should be positive definite. However, for inflationary universes, it must be much larger than one. 

In this paper, we are interested in investigating power-law extensions of the Starobinsky model with a general function of $\hat f(R)$ given by
\begin{equation}
\hat f(R)=R^n,
\end{equation}
where $n$ is an undetermined real number. However, two special values of $n$ such as $n=1$ and $n=2$ will be ignored since they will correspond to the $R$ and $R^2$ terms, respectively. It is interesting that $n=0$ will correspond to a cosmological constant, i.e., $\Lambda =\beta_2 R^0=\beta_2$, and is therefore ignored, too. Note again that several non-trivial values of $n$ can be found in a number of papers mentioned in the introduction, e.g., in Ref.  \cite{Ivanov:2021chn}. 

We will work out the corresponding constraints of $n$ in order to obtain de Sitter solutions with 
\begin{equation}
\alpha(t) =\zeta t, 
\end{equation}
in the context of power-law extensions of the Starobinsky model. Here, $\zeta$ is a parameter, whose value will be solved from the corresponding field equations. Then, we will examine whether the obtained de Sitter solutions are stable or not during an inflationary phase with $\zeta \gg 1$. In order to define field equations, we will use an effective method based on the well-known Euler-Lagrange equations, similar to our previous works \cite{Do:2023yvg,Do:2020vdc,Do:2021fal} as well as other works by other people, e.g., see Ref. \cite{Myrzakulov:2014hca}. Firstly, we will define the corresponding Lagrangian of power-law extension to be
 \begin{equation}\label{key}
\mathcal{L} = \sqrt{-g} \left(R+\beta_1 R^2+\beta_2 R^n \right),
\end{equation}	
where 
\begin{align}
\sqrt{-g} &= N e^{3\alpha},\\
R &=- \dfrac{6}{N^3} \left[ \dot{N}\dot{\alpha}- N \left(2\dot{\alpha}^2+\ddot{\alpha} \right)\right].
\end{align}
Since $\mathcal{L}$ contains the first-order time derivative of $N$ and the second-order time derivative of $\alpha$, the corresponding Euler-Lagrange equations of $N$ and $\alpha$ are defined as
\begin{align}
\dfrac{\partial \Lagr}{\partial N} - \dfrac{d}{dt}\left(\dfrac{\partial \Lagr}{\partial \dot{N}}\right) &= 0,\\
\dfrac{\partial \Lagr}{\partial\alpha} - \dfrac{d}{dt}\left(\dfrac{\partial\Lagr}{\partial \dot{\alpha}}\right)+\dfrac{d^2}{dt^2}\left(\dfrac{\partial\Lagr}{\partial \ddot{\alpha}}\right) &= 0,
\end{align}
respectively. As a result, the explicit forms of these equations turn out to be 	
\begin{align}\label{eq-2.8}
		\dot{\alpha }^2=&-\beta _1 \left(12 \alpha ^{(3)} \dot{\alpha }+36 \dot{\alpha }^2 \ddot{\alpha }-6 \ddot{\alpha }^2\right)+\beta _2 6^{n-1} \left(\ddot{\alpha }+2 \dot{\alpha }^2\right)^{n-2} \nonumber\\
		& \times \left[ -\alpha ^{(3)} \dot{\alpha } (n-1) n + \dot{\alpha }^2 \ddot{\alpha } \left(-4 n^2+7 n-4\right)  +\ddot{\alpha }^2 (n-1) +2 \dot{\alpha }^4 (n-2)\right],\\
		\label{eq-2.9}
		2 \ddot{\alpha }=& -3 \dot{\alpha }^2+\frac{1}{2} \beta _1 \left[-48 \dot{\alpha } \left(\alpha ^{(3)}+4 \dot{\alpha } \ddot{\alpha }\right)+288 \dot{\alpha }^2 \left(\ddot{\alpha }+2 \dot{\alpha }^2\right)-36 \left(\ddot{\alpha }+2 \dot{\alpha }^2\right)^2\right.\nonumber\\
		&\left.+96 \ddot{\alpha } \left(\ddot{\alpha }+2 \dot{\alpha }^2\right)-72 \left(\ddot{\alpha }+2 \dot{\alpha }^2\right) \left(\ddot{\alpha }+3 \dot{\alpha }^2\right)+24 \left(-4 \dot{\alpha } \alpha ^{(3)}-4 \ddot{\alpha }^2-\alpha ^{(4)}\right)\right]\nonumber\\
		&+\frac{1}{2} \beta _2 \left\{-\dot{\alpha } 2^{n+2} 3^{n-1} (1-n) n \left(\alpha ^{(3)}+4 \dot{\alpha } \ddot{\alpha }\right) \left(\ddot{\alpha }+2 \dot{\alpha }^2\right)^{n-2}\right.\nonumber\\
		&\left.+ 2^{n+1} 3^n (1-n) n \dot{\alpha } \left(\alpha ^{(3)}+4 \dot{\alpha } \ddot{\alpha }\right) \left(\ddot{\alpha }+2 \dot{\alpha }^2\right)^{n-2}+ 2^{n+2} 3^n n \dot{\alpha }^2 \left(\ddot{\alpha }+2 \dot{\alpha }^2\right)^{n-1}\right.\nonumber\\
		&\left.+2^{n+2} 3^{n-1} n \ddot{\alpha } \left(\ddot{\alpha }+2 \dot{\alpha }^2\right)^{n-1}-6^n n \left(\ddot{\alpha }+3 \dot{\alpha }^2\right) \left(\ddot{\alpha }+2 \dot{\alpha }^2\right)^{n-1}-6^n \left(\ddot{\alpha }+2 \dot{\alpha }^2\right)^n\right.\nonumber\\
		&\left.-2^{n-2} 3^{n-3} (n-1) n \left(\ddot{\alpha }+2 \dot{\alpha }^2\right)^{n-3} \right. \nonumber\\
		& \left. \times  \left[36 (n-2) \left(\alpha ^{(3)}+4 \dot{\alpha } \ddot{\alpha }\right)^2 -36 \left(-\ddot{\alpha }-2 \dot{\alpha }^2\right) \left(4 \dot{\alpha } \alpha ^{(3)}+4 \ddot{\alpha }^2+\alpha ^{(4)}\right)\right]\right\},	
\end{align}
respectively, after setting $N=1$. Here, $\alpha^{(n)} \equiv d^n\alpha/dt^n$ and so on.  It should be noted that these two field equations are component equations of Einstein field equation. In addition, the second equation \eqref{eq-2.9}, the spatial component of the  Einstein equation, is related to the Friedmann equation \eqref{eq-2.8}, via the Bianchi identity $D_aG^{ab}=0$, where $D_a$ is the covariant derivative. Hence, Eq. \eqref{eq-2.9} can be derived from differentiation and suitable algebraic combinations of  Eq. \eqref{eq-2.8}. Even it is known that Eq. \eqref{eq-2.9} is redundant  \cite{Kao:1999hz}, it is sometimes helpful in solving for the exact solutions. In particular, we will retain these two field equations for the meantime and extract correct set of perturbation equations for our analysis of the corresponding dynamical system of the present model. Indeed, Eq. \eqref{eq-2.9}  will guide us how to define an exact number of suitable dynamical variables. 
\section{Exact de Sitter inflationary solutions} \label{sec3}
 In this section, we are going to figure out de Sitter inflationary solutions  rather than de Sitter non-inflationary solutions. The main reason is that we would like to see if any of considered power-law extensions admits exact de Sitter inflationary solutions, in contrast to the Starobinsky model. However, de Sitter non-inflationary solutions, which could be de Sitter expanding solutions relevant to the late time universe, could also be derived similarly in the present extensions. 

To seek exact de Sitter inflationary solutions, we take the following ansatz for the scale factor as \cite{Barrow:2005qv,Barrow:2006xb,Toporensky:2006kc,Barrow:2009gx,Do:2020vdc,Do:2021fal,Do:2023yvg}
\begin{equation}\label{eq-2.10}
\alpha(t) = \zeta t,
\end{equation}
where $\zeta$ is a parameter, which should be positive definite, i.e., $\zeta >0$, for an expanding universe (or de Sitter non-inflationary universes). For an inflationary phase, however, $\zeta \gg 1$ is necessary. 
As a consequence, Eqs. \eqref{eq-2.8} and \eqref{eq-2.9} both turn into the same equation, 
\begin{equation}\label{eq-2.11}
\beta_2 12^n (n-2) \zeta ^{2n} - 12 \zeta ^2 = 0 ,
\end{equation}
which can be solved to give an exact solution of $\zeta$ such as
\begin{equation}\label{eq-2.12}
\zeta ^{2(n-1)}=\frac{1}{\beta_2 12^{n-1} (n-2)}.
\end{equation}
It is clear that the $R^2$ term does not contribute to the value of $\zeta$, consistent with many previous works, e.g., Ref. \cite{Do:2023yvg}. It is also clear that we will not have solution of $\zeta$ for both $n=1$ and $n=2$. This confirms again an important result that the original Starobinsky model and the pure Einstein gravity do not admit any exact de Sitter solution, e.g., see Refs. \cite{Ketov:2022zhp,Do:2023yvg} for a similar discussion. 

Now, we would like to discuss  the corresponding inflationary solution with $\zeta \gg 1$.  First, its corresponding Hubble parameter is given by
\begin{equation}
H = \dot\alpha =\zeta ,
\end{equation}
which is independent of $\beta_1$ according to Eq. \eqref{eq-2.12} and remains constant during the inflationary phase. While quasi-de Sitter inflationary solutions admit nearly constant Hubble parameters, the constant-like behavior of the Hubble parameter is indeed a unique feature of exact de Sitter inflationary ones,  which would lead to the so-called eternal inflation associated with the multiverse scenario \cite{Guth:2007ng}. 

It is clear that the sign of $\beta_2$ depends on the value of $n$ in order to ensure the existence of real $\zeta$. For example, if $n>2$ then $\beta_2$ must be positive definite. On the other hand, if $n<2$ then $\beta_2$ must be negative definite.  In addition, the absolute value of $\beta_2$ also depends on the value of $n$ in order to fulfill the inflationary constraint $\zeta \gg 1$.  In particular, if $n>1$  then $|\beta_2| \ll 1$ to ensure $\zeta \gg 1$.  On the other hand, if $n<1$ then $|\beta_2| \gg 1$ to ensure $\zeta \gg 1$. The suitable  values of $|\beta_2|$  will, of course, depend on the choice of specific values of $n$. To see how large or small $\beta_2$ should be, we define its expression from Eq. \eqref{eq-2.12} as follows
\begin{equation}\label{2-13}
\beta_2 = \frac{1 }{ 12^{n-1} \zeta ^{2(n-1)} (n-2)}.
\end{equation}
It turns out, for $n>2$, that the increase of $\zeta$ will correspond to the decrease of positive $\beta_2$. Similarly, the larger $\zeta$ is, the smaller $|\beta_2|$ is, for $n\in(1,2)$. In contrast, it appears, for $n<1$, that the increase of $\zeta$ will lead to the increase of $|\beta_2|$. An important question one might ask is that when  $\beta_2$ reaches to its extreme values. This question would be answered if we regard $\beta_2$ as a function of two independent variables $\zeta$ and $n$. Mathematically, the extreme values of $\beta_2$ are solutions of the following set of equations,
\begin{equation}\label{}
	\dfrac{\partial \beta_2}{\partial n} = 0,  \quad \dfrac{\partial \beta_2}{\partial \zeta} = 0.
\end{equation}
As a result,  the first equation, $\partial_n \beta_2 = 0$, gives a non-trivial solution,
	\begin{equation}\label{extreme-sol}
		\zeta^{\text{extr}} = \dfrac{1}{2\sqrt{3}}e^{\frac{1}{2(2-n)}}.
	\end{equation}
The corresponding extremal value of $\beta_2$ can be defined to be
	\begin{equation}\label{}
		\beta^{\text{extr}}_2 = \dfrac{1}{n-2}e^{\frac{n-1}{n-2}},
	\end{equation}
by substituting the solution \eqref{extreme-sol} into Eq. \eqref{2-13}. 
It turns out that $\zeta^{\text{extr}}$ will blow up when $n \to 2^{-}$, i.e.,
\begin{equation}
\lim_{n \to 2^{-}} \zeta^{\text{extr}} = \infty.
\end{equation}
More interestingly, it appears that
\begin{equation}
\lim_{n \to 2^{-}}\beta^{\text{extr}}_2 =0,\quad \lim_{n \to 2^{+}}\beta^{\text{extr}}_2 =\infty.
\end{equation}
Now, we consider the second equation, $\partial_\zeta \beta_2 = 0$. As a result, it can be reduced to  
\begin{equation}\label{}
\dfrac{2(1-n)}{12^{n-1}(n-2)}\zeta^{1-2n}=0.
\end{equation}
It is clear that for $n<1/2$ the left hand side of this equation is always different from zero due to the inflationary constraint $\zeta \gg 1$. This means that the extrema of $\beta_2$ will not exist for $n<1/2$. On the other hand, for $n>1/2$ the extrema of $\beta_2$ could appear at $\zeta = \infty$. Combining these analysis, we arrive at a final conclusion that $\beta_2$ tends to its extrema  when $n \to 2^{-}$ along with $\zeta \to \infty$.

In the following subsections, we will consider four ranges of $n$ as follows: (i) $n<0$; (ii) $0<n<1$; (iii) $1<n<2$; and (iv) $n>2$. 
\subsection{Range 1: $n<0$}
\begin{figure}[htp!]
			\centering
				\includegraphics[scale=0.9]{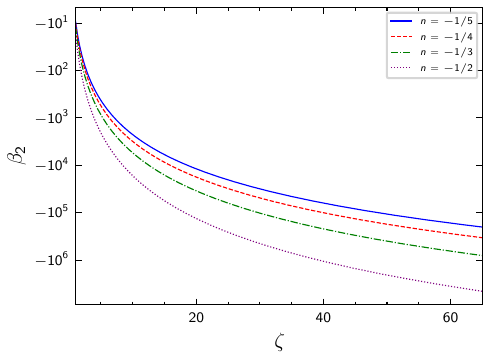} \quad
				\includegraphics[scale=0.5]{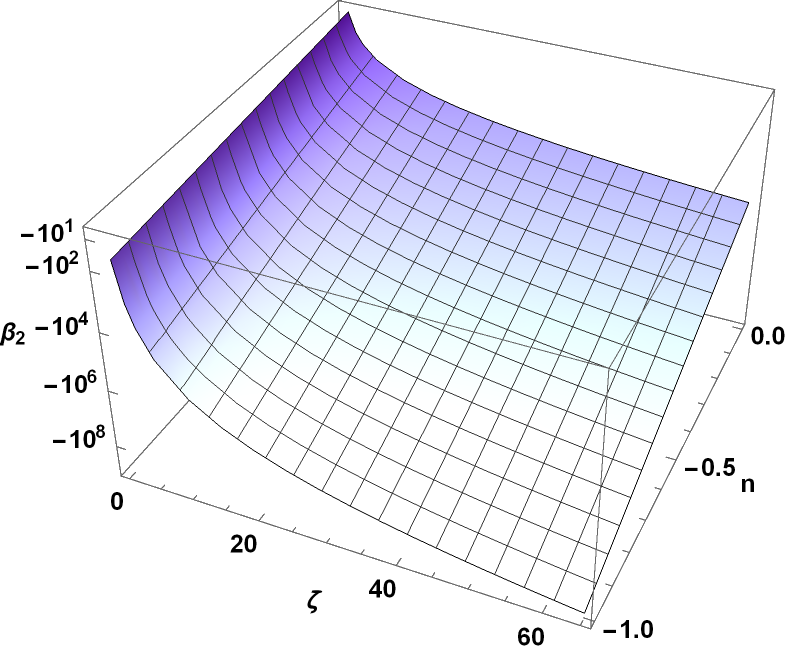}\\
			\caption{(Left figure) $\beta_2$ as a function of $\zeta$ for different values of $n\in(-1,0)$. (Right figure) $\beta_2$ as a function $n\in(0,-1)$ and $\zeta\in(1,65)$.}
			\label{fig:n-0}
\end{figure}

\begin{figure}[htp!]
	\centering
		\includegraphics[scale=0.9]{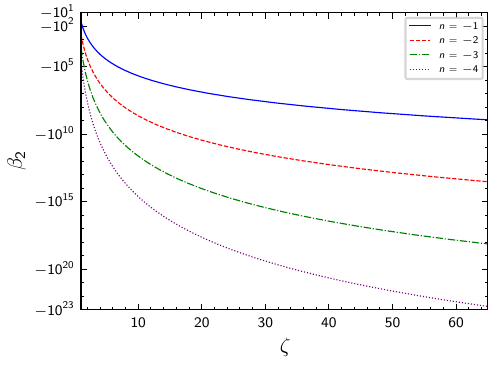}\quad
		\includegraphics[scale=0.5]{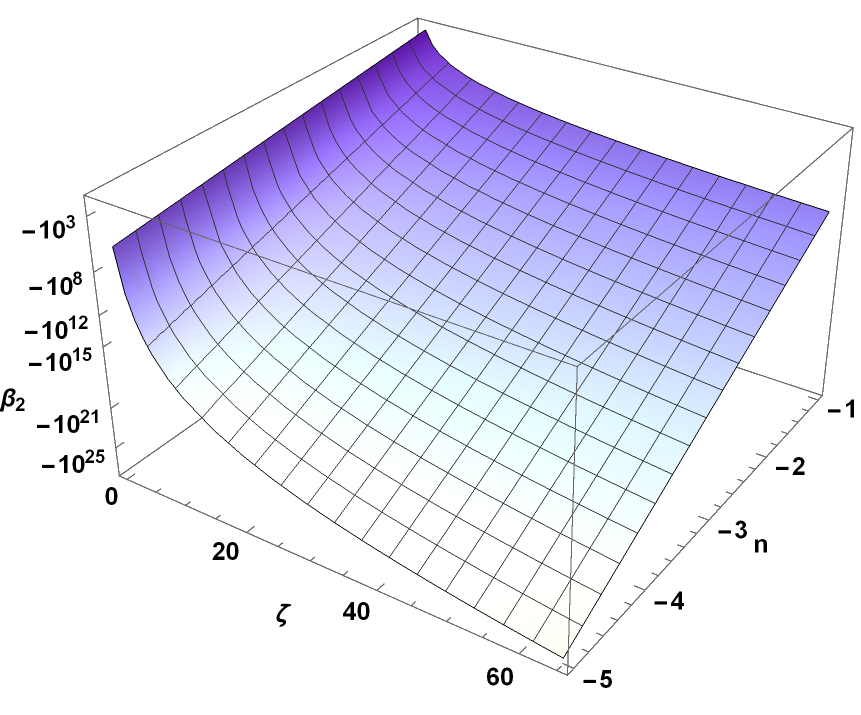}\\
		\caption{(Left figure) $\beta_2 $ as a function of $\zeta $ for different values of $n\leq -1$. (Right figure) $\beta_2 $ as a function of $n\in (-5,-1)$ and $\zeta\in(1,65)$.}
	\label{fig:3D-n-0}
\end{figure}

In this range, to ensure the existence of  inflation with $\zeta\gg 1$, $|\beta_2|$ must be much larger one. For example, we list here several values of $n$ to figure out the corresponding value of $\beta_2$ as follows,
\begin{align}
			&\text{for}\quad n=-\frac{1}{2} \quad\text{then}\quad \beta_2 \simeq -3.59\times 10^6,\\
			&\text{for}\quad n=-1 \quad\text{then}\quad \beta_2 \simeq -6.22\times 10^8,\\
			&\text{for}\quad n=-2 \quad\text{then}\quad \beta_2 \simeq -2.02\times 10^{13},\\
			&\text{for}\quad n=-3 \quad\text{then}\quad \beta_2 \simeq  -6.97\times 10^{17},
\end{align}
provided that $\zeta =60$. For more details, see Fig. \ref{fig:n-0}. In particular, the left figure is plotted for the parameter $\beta_2$ acting as a function of the $\zeta$ in the specific range, $-1<n<0$, while the right figure is plotted for the parameter $\beta_2$ acting as a function of $n$ and $\zeta$ with chosen ranges are $n\in (-1,0)$ and $\zeta\in(1,65)$. Additional details for $n\leq -1$ can be seen in Fig. \ref{fig:3D-n-0}.
\subsection{Range 2: $0<n<1$} 
\begin{figure}[htp!]
		\centering
			\includegraphics[scale=0.9]{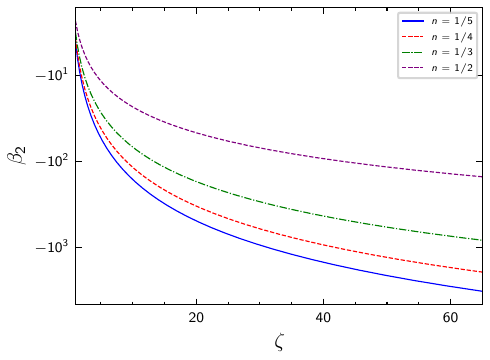}\quad
			\includegraphics[scale=0.5]{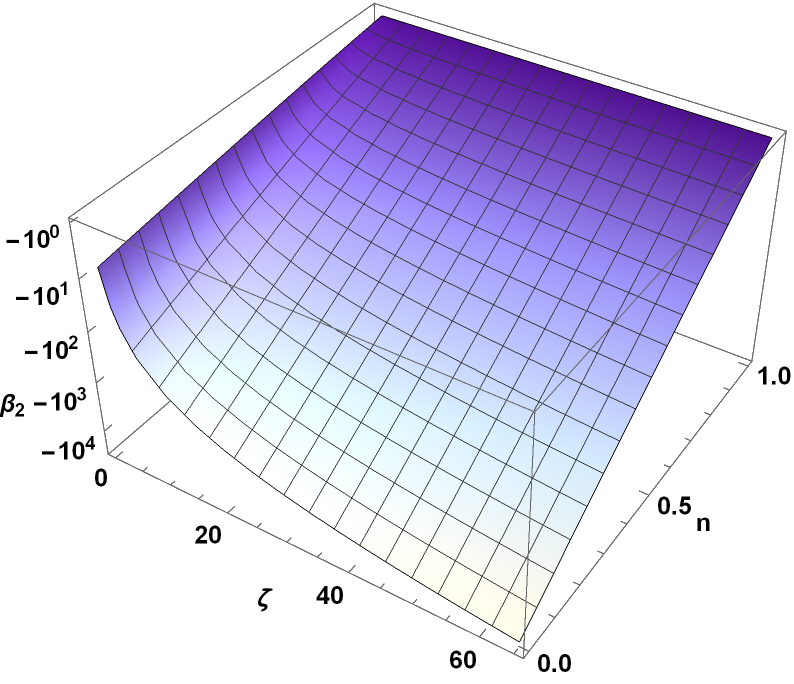}\\
		\caption{(Left figure) $\beta_2$ as a function of $\zeta$ for different values of {$n \in (0,1)$}. (Right figure) $\beta_2$ as a function of $n\in(0,1)$ and $\zeta\in (1,65)$.}
		\label{fig:3D-n-0-1}
	\end{figure}

In this range, it turns out that $-6 \zeta^2<\beta_2<-1$.  For details, see Fig. \ref{fig:3D-n-0-1} with two figures, which are plotted in the range $ 0<n<1$. According to these figures, it appears that $|\beta_2|\gg 1$ is the necessary condition  to ensure $\zeta\gg 1$.
\subsection{Range 3: $1<n<2$}
\begin{figure}[htp!]
		\centering
			\includegraphics[scale=0.9]{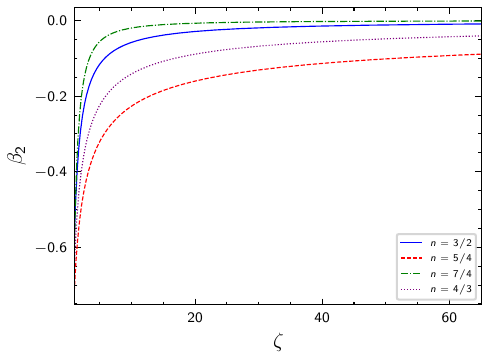} \quad
			\includegraphics[scale=0.5]{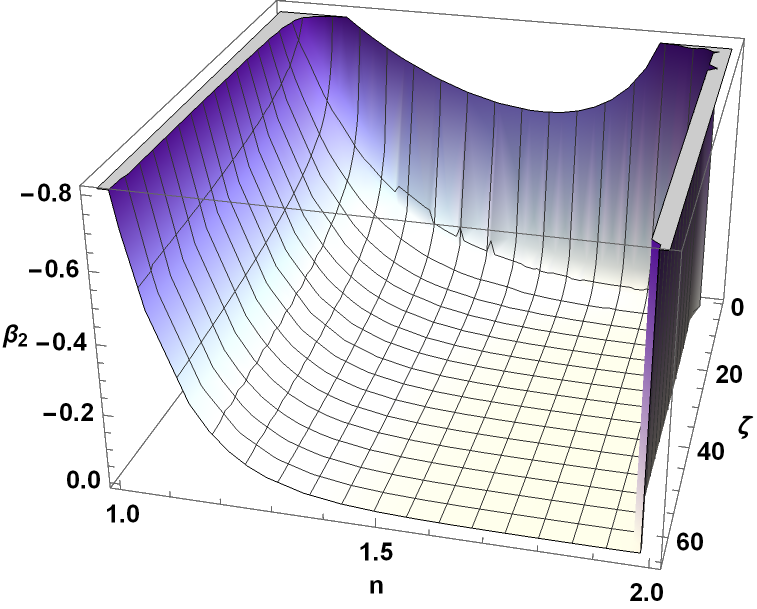}\\
		\caption{(Left figure) $\beta_2$ as a function of $\zeta$ for different values of {$n \in (1,2)$}. (Right figure) $\beta_2$ as a function of  $n\in(1,2)$ and $\zeta\in (1,65)$.}
		\label{fig:3D-n-1-2}
	\end{figure}

In this range, it turns out that the following constraint $-1<\beta_2<0$ is needed. Of course, the suitable value of $\beta_2$ to guarantee the existence of inflation with $\zeta\gg1$ depends on the specific value of $n$.  As a specific example, the choice $n=3/2$ and $\zeta = 60$ will lead to the corresponding value $\beta_2 \simeq -9.62 \times 10^{-3}$, according to Eq. \eqref{2-13}.  See Fig. \ref{fig:3D-n-1-2} for other choices and for the relationships among these parameters. 
\subsection{Range 4: $n>2$}
\begin{figure}[htp!]
	\centering
		\includegraphics[scale=0.9]{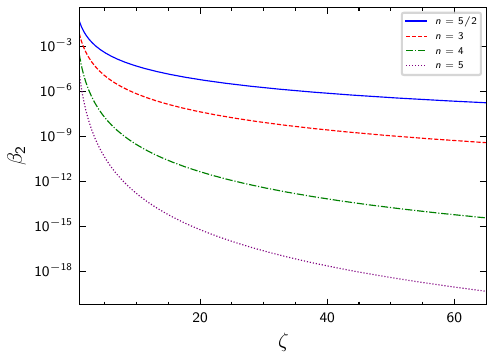} \quad
		\includegraphics[scale=0.5]{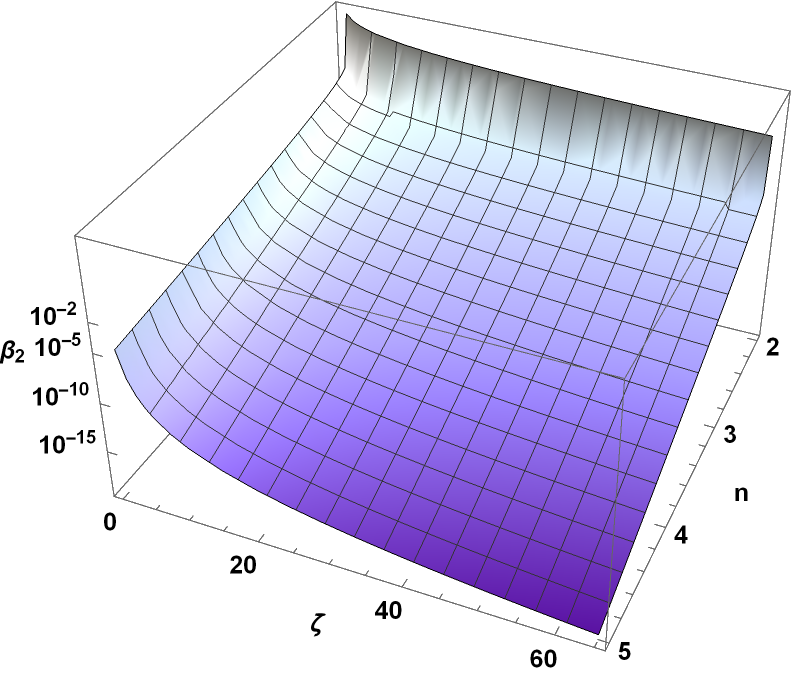}\\
	\caption{(Left figure) $\beta_2$ as a function of $\zeta$ for different values of $n>2$. (Right figure) $\beta_2 $ as a function of $n>2$ and $\zeta\in (1,65)$.}
	\label{fig:3D-n-2-5}
\end{figure}

 In this range, $0<\beta_2\ll 1$ is needed according to Eqs. \eqref{eq-2.12} and \eqref{2-13}. More concretely, we also take some explicit examples for fixed value $\zeta=60$ as follows
	\begin{align}
		&\text{for}\quad n=3 \quad\text{then}\quad \beta_2 \simeq 5.36\times 10^{-10},\\
		&\text{for}\quad n=4\quad\text{then}\quad \beta_2 \simeq  6.20\times 10^{-15}.
	\end{align}
It is straightforward to see that the value for $\beta_2$ is positive and much smaller than one to ensure $\zeta\gg1$. See Fig. \ref{fig:3D-n-2-5} for the relationships among $\beta_2$, $n$, and $\zeta$.
	
	In summary, we have considered four different ranges of values of $n$ and accordingly examined the values of $\beta_2$ to get the exact  de Sitter inflationary solutions $\zeta\gg1$. Apparently, the parameter $\beta_2$ must be positive or negative definite for $n<2$ or $n>2 $, respectively. Keeping in mind this useful information, we are going to investigate the stability of de Sitter inflationary solutions in the next section.
\section{Dynamical system and its fixed point}\label{sec4}
In this section, we would like to investigate whether the obtained de Sitter inflationary solutions are stable or not. For this purpose, we will transform the field equations into the corresponding dynamical system, similar to the previous works \cite{Barrow:2005qv,Barrow:2006xb,Barrow:2009gx,Muller:2017nxg,Do:2020vdc,Do:2021fal,Do:2023yvg}, by introducing dimensionless dynamical variables such as
	\begin{align}
		B= \dfrac{1}{\dot{\alpha}^2},\quad Q = \dfrac{\ddot{\alpha}}{\dot{\alpha}^2},\quad Q_2 = \dfrac{\alpha^{(3)}}{\dot{\alpha}^3},
	\end{align}
here  ${\dot{\alpha}\equiv H} $ is nothing but the Hubble parameter. As a result,  the corresponding set of autonomous equations of dynamical variables are defined to be
	\begin{align} \label{dynamical-1}
		B' &= -2 B Q,\\
		 \label{dynamical-2}
		Q' &= Q_2 -2 Q^2,\\
		 \label{dynamical-3}
		Q'_2 &=\alpha^{(4)} B^2 - 3 Q Q_2,
	\end{align}	
where $B' \equiv dB/d\tau$, $Q'\equiv dQ/d\tau$, and $Q'_2 \equiv dQ_2/d\tau$, with $\tau \equiv \int \dot\alpha dt$ introduced as a dynamical time variable. Furthermore, the field equations \eqref{eq-2.8} and \eqref{eq-2.9} can be rewritten in terms of the dynamical variables as 
	\begin{align}\label{eq-3.7}
		&6 B (Q+2)+\beta _1 \left(-36 Q^3+144 Q^2+72 Q_2 Q+432 Q+144 Q_2\right)\nonumber\\
		&+\beta _2  6^n B \left(\frac{Q+2}{B}\right)^{n-1} \left(4 n^2 Q+n^2 Q_2-n Q^2-7 n Q-n Q_2-2 n+Q^2+4 Q+4\right) =0, \\
		\label{eq-3.8}		
		&6 B (Q+2)^2 (2 Q+3)+\beta _1 \left[72 \alpha ^{(4)} B^2 (Q+2)^2+324 Q (Q+2)^3+432 Q_2 (Q+2)^2\right]\nonumber\\
		&+\beta _2  6^n B \left(\frac{Q+2}{B}\right)^{n-1} \left[Q \left(\alpha ^{(4)} B^2 n^2-\alpha ^{(4)} B^2 n+8 n^3 Q_2-18 n^2 Q_2+16 n^2 \right. \right. \nonumber\\
		& \left. \left. +10 n Q_2-32 n+36\right)+(n-1) n \left(2 \alpha ^{(4)} B^2+n Q_2^2-2 Q_2^2+12 Q_2\right)+\left(4 n^2-5 n+3\right) Q^3 \right. \nonumber\\
		&\left. +\left(16 n^3-32 n^2+9 n+18\right) Q^2-12 (n-2)\right]=0,
	\end{align}	
respectively. It should be noted that the expression of $\alpha^{(4)}$ in terms of dynamical variables can be derived from Eq. \eqref{eq-3.8}. In addition, Eq. \eqref{eq-3.7} coming from the Friedmann equation \eqref{eq-2.8} will act as a constraint equation that any fixed points of the dynamical system should be satisfied. 

Given the above dynamical system, it is easy to find the corresponding fixed points, which are nothing but solutions of the following set of equations, 
	\begin{equation}\label{}
		B' = Q'  = Q'_2 = 0.
	\end{equation}
As a result, it turns out, according to Eqs.  \eqref{dynamical-1}, \eqref{dynamical-2}, and \eqref{dynamical-3}, that
	\begin{equation}\label{20}
		Q=Q_2 = \alpha^{(4)}= 0.
	\end{equation}
Substituting these results into both Eqs. \eqref{eq-3.7} and \eqref{eq-3.8}, we obtain the same equation,
	\begin{align}\label{eq-3.9}
		12^{n-1} \beta_2 (n-2) B^{1-n}=1,
	\end{align}
which is really consistent with Eq. \eqref{eq-2.11}. Indeed, integrating out this equation leads to an exact solution,
	\begin{equation}\label{}
		\alpha(t) = \zeta t,
	\end{equation}
which is nothing but the ansatz shown in Eq. \eqref{eq-2.10} with $\zeta$ defined in Eq. \eqref{eq-2.12}. This implies that the fixed point of dynamical system is completely equivalent to the de Sitter solution of field equations  found in the previous section. Therefore, the stability of the fixed point during the inflationary phase will tell us the stability of the corresponding de Sitter inflationary  solution.
\section{Stability analysis} \label{sec5}
In this {section}, we will examine the stability of the fixed point. To do this task, we first perturb the autonomous equations such as
	\begin{align}
		\delta B'&=-2 B \delta Q-2 Q\delta B ,\\
		\delta Q'&=\delta Q_2-4Q\delta Q ,\\
		\delta Q_2'&={\delta \alpha ^{(4)} B^2+2\alpha ^{(4)} B \delta B}-3 Q\delta Q_2 -3Q_2 \delta Q.
	\end{align}
Considering perturbations around the fixed point with $Q=Q_2 = \alpha^{(4)}= 0$, these perturbed equations reduce to
	\begin{align}\label{26}
		\delta B' &=-2 B\delta Q,\\\label{27}
		\delta Q' &= \delta Q_2,\\\label{28}
		\delta Q'_2 &= {\delta\alpha^{(4)} B^2.}
	\end{align}
To determine $\delta\alpha^{(4)}$, we perturb Eqs. \eqref{eq-3.7} and \eqref{eq-3.8} around the fixed point as follows
	\begin{align}\label{29}
		& \left[432 \beta _1+\beta _2 3^n 4^{n-1} \left(6 n^2-8 n+4\right) B^{2-n}+6 B\right] \delta Q + \left[\beta _2 12^n (n-2)^2 B^{1-n}+12\right] \delta B\nonumber\\
		&+\left[144 \beta _1+\beta _2 2^{2 n-1} 3^n (n-1) n B^{2-n}\right] \delta Q_2  =0,\\\label{30}
		& B^2 \left[288 \beta _1+\beta _2 12^n (n-1) n B^{2-n}\right] \delta\alpha^{(4)} \nonumber\\
		& +\frac{1}{4}   \left[10368 \beta _1+\beta _2 3^n 4^{n+1} \left(5 n^2-7 n+12\right) B^{2-n}+480 B\right]\delta Q \nonumber\\
		&+\frac{1}{2}  \left[\beta _2 12^{n+1} (n-2)^2 B^{1-n}+144\right] \delta B+ \left[1728 \beta _1+\beta _2 2^{2 n+1} 3^{n+1} (n-1) n B^{2-n}\right] \delta Q_2=0.
	\end{align}
As a result, we obtain  from Eq. \eqref{29} that
	\begin{equation}\label{}
		\delta B= -\frac{ 864 \beta _1+\beta _2 12^n \left(3 n^2-4 n+2\right) B^{2-n}+12 B}{ 2\beta _2 12^n (n-2)^2 B^{1-n}+24}\delta Q-\frac{288 \beta _1+\beta _2 12^n (n-1) n B^{{2-n}}}{2\beta _2 12^n (n-2)^2 B^{1-n}+24}\delta Q_2.
	\end{equation}
Furthermore, substituting this expression into Eq. \eqref{30} gives
	\begin{equation}\label{}
		\delta\alpha^{(4)}=\frac{\beta _2 12^n (n-2) (4 n+3) B^{1-n}-84}{288 \beta _1 B+\beta _2 12^n (n-1) n B^{3-n}}\delta Q-\frac{3 }{B^2}\delta Q_2.
	\end{equation}
As we mentioned earlier in the section \ref{sec2},  we take advantage of the helpful combination of field equations \eqref{29} and \eqref{30} to express the perturbation $\delta \alpha^{(4)}$ for convenience.

Now, we are going to take exponential perturbations,
	\begin{align}\label{}
		\delta B= A_B \exp[\lambda\tau],\\
		\delta Q = A_Q \exp [\lambda \tau],\\
		\delta Q_2 = A_{Q_2}\exp[\lambda \tau],
	\end{align}
where $A_B$, $A_Q$, and $A_{Q_2}$ are undetermined constants. It is noted that the sign of $\lambda$ will judge the stability of fixed point.  In particular, if $\lambda$ is found to be positive definite then all perturbations will blow up when $\tau$ becomes large. In other words, the fixed point will be unstable if $\lambda >0$. Otherwise, the fixed point will be stable if $\lambda <0$ since all perturbations will tend to be vanished as $\tau \to \infty$.  As a result, the perturbed equations \eqref{26}, \eqref{27}, and \eqref{28} are reduced to the following homogeneous set of algebraic equations of $A_B$, $A_Q$, and $A_{Q_2}$,
	\begin{align}
		-\lambda A_B   -2 B  A_Q&=0,\\
		\lambda A_Q  - A_{Q_2}  &= 0,\\
		{\frac{\beta _2 12^n (n-2) B (4 n+3) B^{1-n}-84B}{288 \beta _1 +\beta _2 12^n (n-1) n B^{2-n}}}A_Q -\left(\lambda+ 3\right)A_{Q_2}&= 0.
	\end{align}
For convenience, this set will be written in the following matrix equation such as
\begin{align}\label{}
		\mathcal{M}  \begin{pmatrix}
			A_B\\
			A_Q\\
			A_{Q_2}
		\end{pmatrix}&\equiv	
		\begin{bmatrix}
			M_{11}  & M_{12} & 0 \\
			0 & M_{22}  & M_{23}  \\
			0 & M_{32} & M_{33} \\
		\end{bmatrix}
		\begin{pmatrix}
			A_B\\
			A_Q\\
			A_{Q_2}
		\end{pmatrix}=0,
	\end{align}
where non-vanishing elements of matrix $\mathcal{M}$ are given by
	\begin{widetext}
	\begin{align}
		M_{11} &= -\lambda,\quad M_{12} = -2 B,\\
		 M_{22} &= -\lambda,\quad M_{23} =1,\\
		 M_{32} & ={\frac{\beta _2 12^n (n-2) B (4 n+3) B^{1-n}-84B}{288 \beta _1 +\beta _2 12^n (n-1) n B^{2-n}}} ,\quad M_{33} =-\lambda-3. 
	\end{align}
	\end{widetext}
Mathematically, this matrix equation has non-trivial solutions if and only if
	\begin{equation}
		\det \mathcal{M} = 0,
	\end{equation}
which can be expanded as an analytic equation for $\lambda$ such as
	\begin{equation}\label{poly-eq}
	\lambda\left(a_3 \lambda^2 +a_2\lambda+a_1 \right) =0,
	\end{equation}
	where
	\begin{align}
		a_3 &= -1 <0,\\\label{eq-3.34}
		a_2 &= -3<0,\\\label{eq-3.35}
		a_1 &=  {\frac{\beta _2 12^n (n-2) B (4 n+3) B^{1-n}-84B}{288 \beta _1 +\beta _2 12^n (n-1) n B^{2-n}}}.
	\end{align}
For the de Sitter inflationary solution found in the previous section with $\zeta\gg1 $, it is clear that $0<B=1/\zeta^2\ll 1$.  Apparently, the stability of fixed point \eqref{eq-3.9} now depends only on the sign of coefficient $a_1$. In particular, if $a_1\leq0$ then the fixed point will be stable since the equation \eqref{poly-eq} will not admit any positive root $\lambda >0$. On the other side, if $a_1> 0$ then the fixed point will become unstable since the equation \eqref{poly-eq} will admit at least one positive root $\lambda>0$. In this paper, we will consider both cases, $\beta_2>0$ and $\beta_2<0$. Furthermore,  $\beta_1$ will be regarded as a free parameter, similar to our previous work \cite{Do:2023yvg}. 

It should be noted that in the following analysis, we simplify the coefficient $a_1$ from Eq. \eqref{eq-3.35} as
\begin{equation}\label{eq-3.39}
	a_1 = \frac{4 B (n-2) (n-1)}{B (n-1) n+24 \beta_1  (n-2)},
\end{equation}
thanks to a relation,
\begin{equation}\label{}
	B^{1-n} = \dfrac{1}{12^{n-1}\beta_2(n-2)},
\end{equation}
which is derived from Eq.  \eqref{eq-3.9}. In a special case of $\beta_1=0$, the exact de Sitter solution \eqref{eq-2.12} remains unchanged due to the fact that Eq \eqref{eq-2.11} does not include $\beta_1$. However, $a_1$ will reduce to 
\begin{equation}
	a_1 = \dfrac{4(n-2)}{n},
	\end{equation}
which implies that the corresponding de Sitter solution will be stable if and only if $0<n<2$, or equivalently $a_1<0$. Otherwise, it will be unstable accordingly.

The expression shown in Eq. \eqref{eq-3.39}  implies that the positivity or negativity of $a_1$ does not depend on $\beta_2$ explicitly but depends on the sign of $\beta_1 \neq 0$ and $B$. Let us discuss them for now.
\subsection{Case of negative $\beta_2$}
As discussed above, we will consider three different ranges of $n$ as follows: (i) $n<0$, (ii) $0<n<1$, and (iii) $1<n<2$. Additionally, we will also examine both $\beta_1>0$ and $\beta_1<0$.
\subsubsection{The range $n<0$}
In this range, the corresponding de Sitter inflationary solution will appear when {$\beta_2<0$ and  $|\beta_2|\gg 1$}. 

$\bullet$ $\beta_1>0$: According to Eq. \eqref{eq-3.39}, the {numerator} of $a_1$ is consistently positive, leading to the sign of $a_1$ being determined solely by the {denominator}. Consequently, in order to obtain a stable fixed point, i.e., $a_1<0$ , the condition 
	\begin{equation}\label{}
		{B (n-1) n+24\beta_1  (n-2) <0,}
	\end{equation}
should be satisfied. This condition is equivalent to
	\begin{equation}\label{}
		B<-\frac{24 \beta_1  (n-2)}{(n-1) n}.
	\end{equation}
If this equation holds true, then the fixed point in this particular case turns into stability.

$\bullet$ $\beta_1<0$: On the contrary, one can easily verify that when $\beta_1<0$, the sign of the coefficient $a_1$ always becomes positive, this result implies that the fixed point is unstable in this scenario.
\subsubsection{The range $0<n<1$}
In this range, the corresponding de Sitter inflationary solution will be guaranteed to exist if the constraint $-6\zeta^2<\beta_2<-1$ is met.

$\bullet$ $\beta_1>0$: Firstly, we need to note that the sign of $a_1$ is consistently negative as indicated by Eq. \eqref{eq-3.39} due to its {numerator} being positive while the {denominator} tends to be negative. Consequently, this implication leads to the fact that the fixed point will always remain stable.

$\bullet$ $\beta_1<0$: Furthermore, when $\beta_1<0$, the following constraint,
	\begin{equation}\label{eq-4.44}
		B>-\frac{24 \beta_1  (n-2)}{(n-1) n},
	\end{equation}
is required to ensure the stability of the fixed point. 
\subsubsection{The range $1<n<2$}
In this range,  the following constraint $-1<\beta_2<0$ must be satisfied to ensure the existence of the corresponding de Sitter inflationary solution. 

$\bullet$ $\beta_1>0$: Due to the inequalities $\beta_1>0$, $1<n<2$, and $B>0$, the sign of $a_1$  depends only on the sign of the denominator in Eq. \eqref{eq-3.39} since its numerator is always negative. Therefore, the inequality,
	\begin{equation}\label{}
		{B (n-1) n+24 \beta_1  (n-2) >0,}
	\end{equation}
needs to be fulfilled in order to ensure $a_1<0$. Consequently, the following constraint for $B$ as
	\begin{equation}\label{eq-4.43}
		B>-\frac{24 \beta_1  (n-2)}{(n-1) n},
	\end{equation}
is necessary for the stability of the fixed point.

$\bullet$ $\beta_1<0$: It appears that the sign of coefficient $a_1$ in Eq. \eqref{eq-3.39} is always negative due to the fact that its {numerator} is negative while its {denominator} is positive. Hence, the fixed point is always stable.
\subsection{Case of positive $\beta_2$}
In this case,  $n$ must be larger than two in order to ensure the existence of the corresponding de Sitter inflationary solution with $\zeta\gg 1$ as pointed out in the previous section.

$\bullet$ $\beta_1>0$:  It appears that $a_1$ is always positive definite in this case since $n>2$ and $B>0$. Hence, the fixed point will  always be unstable.

$\bullet$ $\beta_1<0$:  According to Eq. \eqref{eq-3.39}, it is clear that the sign of $a_1$ depends only on the sign of the denominator because the numerator is always positive for $n>2$. In particular, the negativity of $a_1$ will require the following inequality,
	\begin{equation}
		{B (n-1) n+24 \beta_1  (n-2) <0,}
	\end{equation}
or equivalently,
	\begin{equation}\label{inequality-1}
		B<-\frac{24 \beta_1  (n-2)}{(n-1) n}.
	\end{equation}
Of course, the fixed point will be stable for $\beta_1<0$ if the inequality \eqref{inequality-1} happens during the inflationary phase. Otherwise, the fixed point will be unstable.

	\begin{widetext}
		For convenience, the stability analysis of the de Sitter inflationary solution will be summarized  in the Table \ref{tab:my-table}. 
	\begin{center}
	\begin{table}[htp!]
		\begin{tabular}{@{}cclclcccccc@{}}
			\hline
			\multicolumn{1}{|c|}{Ranges of $n$} &
			\multicolumn{4}{c|}{$n>2$} &
			\multicolumn{2}{c|}{${1<n<2}$} &
			\multicolumn{2}{c|}{$0<n<1$} &
			\multicolumn{2}{c|}{$n<0$} \\ 
			\hline
			\multicolumn{1}{|c|}{Sign of $\beta_2$} &
			\multicolumn{4}{c|}{$\beta_2>0$} &
			\multicolumn{6}{c|}{$\beta_2<0$} \\ 
			\hline
			\multicolumn{1}{|c|}{Sign of $\beta_1$} &
			\multicolumn{2}{c|}{$\beta_1>0$} &
			\multicolumn{2}{c|}{$\beta_1<0$} &
			\multicolumn{1}{c|}{$\beta_1>0$} &
			\multicolumn{1}{c|}{$\beta_1<0$} &
			\multicolumn{1}{c|}{$\beta_1>0$} &
			\multicolumn{1}{c|}{$\beta_1<0$} &
			\multicolumn{1}{c|}{$\beta_1>0$} &
			\multicolumn{1}{c|}{$\beta_1<0$} \\ 
			\hline
			\multicolumn{1}{|c|}{\multirow{4}{*}{$B<-\frac{24 \beta_1  (n-2)}{(n-1) n}$}} &
			\multicolumn{2}{c|}{\multirow{8}{*}{saddle}} &
			\multicolumn{2}{c|}{\multirow{4}{*}{attractor}} &
			\multicolumn{1}{c|}{\multirow{4}{*}{saddle}} &
			\multicolumn{1}{c|}{\multirow{8}{*}{attractor}} &
			\multicolumn{1}{c|}{\multirow{8}{*}{attractor}} &
			\multicolumn{1}{c|}{\multirow{4}{*}{saddle}} &
			\multicolumn{1}{c|}{\multirow{4}{*}{attractor}} &
			\multicolumn{1}{c|}{\multirow{8}{*}{saddle}} \\
			\multicolumn{1}{|c|}{} &
			\multicolumn{2}{c|}{} &
			\multicolumn{2}{c|}{} &
			\multicolumn{1}{c|}{} &
			\multicolumn{1}{c|}{} &
			\multicolumn{1}{c|}{} &
			\multicolumn{1}{c|}{} &
			\multicolumn{1}{c|}{} &
			\multicolumn{1}{c|}{} \\
			\multicolumn{1}{|c|}{} &
			\multicolumn{2}{c|}{} &
			\multicolumn{2}{c|}{} &
			\multicolumn{1}{c|}{} &
			\multicolumn{1}{c|}{} &
			\multicolumn{1}{c|}{} &
			\multicolumn{1}{c|}{} &
			\multicolumn{1}{c|}{} &
			\multicolumn{1}{c|}{} \\
			\multicolumn{1}{|c|}{} &
			\multicolumn{2}{c|}{} &
			\multicolumn{2}{c|}{} &
			\multicolumn{1}{c|}{} &
			\multicolumn{1}{c|}{} &
			\multicolumn{1}{c|}{} &
			\multicolumn{1}{c|}{} &
			\multicolumn{1}{c|}{} &
			\multicolumn{1}{c|}{} \\ \cmidrule(r){1-1} \cmidrule(lr){4-6} \cmidrule(lr){9-10}
			\multicolumn{1}{|c|}{\multirow{4}{*}{$B>-\frac{24 \beta_1  (n-2)}{(n-1) n}$}} &
			\multicolumn{2}{c|}{} &
			\multicolumn{2}{c|}{\multirow{4}{*}{saddle}} &
			\multicolumn{1}{c|}{\multirow{4}{*}{attractor}} &
			\multicolumn{1}{c|}{} &
			\multicolumn{1}{c|}{} &
			\multicolumn{1}{c|}{\multirow{4}{*}{attractor}} &
			\multicolumn{1}{c|}{\multirow{4}{*}{saddle}} &
			\multicolumn{1}{c|}{} \\
			\multicolumn{1}{|c|}{} &
			\multicolumn{2}{c|}{} &
			\multicolumn{2}{c|}{} &
			\multicolumn{1}{c|}{} &
			\multicolumn{1}{c|}{} &
			\multicolumn{1}{c|}{} &
			\multicolumn{1}{c|}{} &
			\multicolumn{1}{c|}{} &
			\multicolumn{1}{c|}{} \\
			\multicolumn{1}{|c|}{} &
			\multicolumn{2}{c|}{} &
			\multicolumn{2}{c|}{} &
			\multicolumn{1}{c|}{} &
			\multicolumn{1}{c|}{} &
			\multicolumn{1}{c|}{} &
			\multicolumn{1}{c|}{} &
			\multicolumn{1}{c|}{} &
			\multicolumn{1}{c|}{} \\
			\multicolumn{1}{|c|}{} &
			\multicolumn{2}{c|}{} &
			\multicolumn{2}{c|}{} &
			\multicolumn{1}{c|}{} &
			\multicolumn{1}{c|}{} &
			\multicolumn{1}{c|}{} &
			\multicolumn{1}{c|}{} &
			\multicolumn{1}{c|}{} &
			\multicolumn{1}{c|}{} \\ 
			\hline
		\end{tabular}
		\caption{The summary of stability analysis of the de Sitter inflationary solution for different ranges of $n$, $\beta_1$, and $\beta_2$.}
		\label{tab:my-table}
	\end{table}
\end{center}
	\end{widetext}
\section{Some specific models} \label{sec6}
In this section, we will use numerical methods to illustrate the stability analysis, focusing on specific models within distinct ranges of the parameter $n$. Note again that some values of $n$ such as $n=3/2$ and $3$ have been considered in a recent paper \cite{Ivanov:2021chn}. To do this task, we will investigate the following models:  For the range $n>2$, we choose a model with $n=3$; for the range $1<n<2$, we choose a model with $n=3/2$; for the range $0<n<1$, we choose a model with $n=1/2$; for the range $n<0$, we choose a model with $n=-1$.
\subsection{$\hat{f}(R) = R^3$ model} \label{subsection-R3}
For the model with $n=3$, as discussed in the previous section, there exists a  region, where the de Sitter fixed point becomes stable during the inflationary phase, characterized by $\beta_1<0$ and $ B<-\frac{24 \beta_1  (n-2)}{(n-1) n} $ as indicated in the Table \ref{tab:my-table}. As a consequence,  the corresponding constraint for the stability of the fixed point is given by
\begin{equation}\label{}
	0<\beta_2 <\frac{\beta_1 ^2}{9}.
\end{equation}
For an explicit value for $\beta_1 = -10^{-4}$, the corresponding inequality for $\beta_2$ turns out to be
\begin{equation}\label{5.2}
	\beta_2 < 1.11\times 10^{-9}.
\end{equation}
Therefore, we can choose a value for $\beta_2$ such as $\beta_2 = {\beta_1^2}/{12} \simeq 8.33\times 10^{-10}$, which fulfills the condition \eqref{5.2}, for numerical calculations. Consequently, the corresponding approximated value of $a_1$ reads
\begin{equation}\label{}
	a_1\simeq{-8.62} <0.
\end{equation}
 \begin{figure}[hbtp] 
 \begin{center}
 	\includegraphics[scale=0.5]{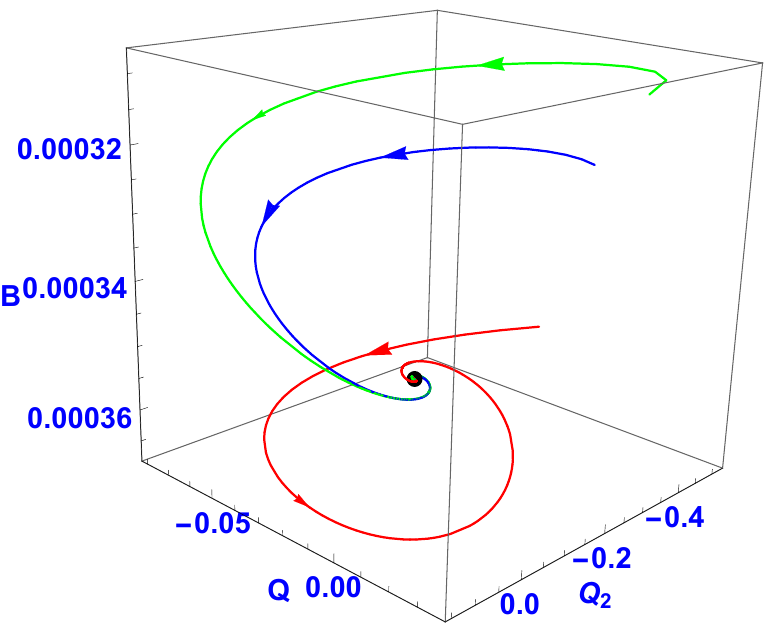}\\
 	\caption{Attractive behavior of the de Sitter fixed point of the $\hat{f}(R) = R^{3}$ model (displayed as a {black point}). Here, the parameters have been chosen as $\beta_1 = -10^{-4}$ and $\beta_2= \beta_1^2/12$. 
	}
 	\label{fig:numerical-n-2}
	\end{center}
 \end{figure}
It becomes clear that numerical results confirm our stability analysis made above. To be more specific, the fixed point (displayed as a {black point}), which is in compatible with the inequalities shown above, does act as an attractor point as expected as shown in Fig. \ref{fig:numerical-n-2}.  Thus, the corresponding de Sitter inflationary solution of this model becomes stable. Furthermore, by doing the same analysis we can be able to show that any model with  $n>2$ and $n\neq 3$ will also admit its stable de Sitter inflationary solution. 
\subsection{$\hat{f}(R) = R^{3/2}$ model}
We can now verify the stability constraints by examining another specific model with $n=3/2$, as discussed in Ref. \cite{Ivanov:2021chn}. As pointed out in Table \ref{tab:my-table}, there exist two distinct regions where the de Sitter inflationary solution becomes stable.
The first region corresponds to $\beta_1>0$ along with
\begin{equation}\label{5.3}
	B>-\frac{24 \beta_1  (n-2)}{(n-1) n},
\end{equation}
while the second one is associated with $\beta_1<0$. 
\subsubsection{First region with $\beta_1>0$} 
\begin{figure}[hbtp] 
	\centering
	\includegraphics[scale=0.5]{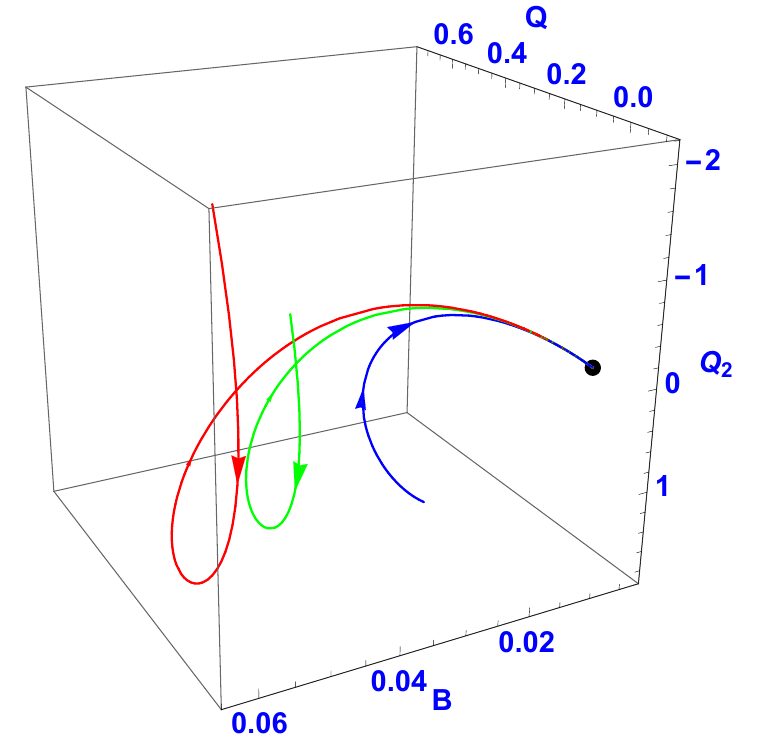}
	\caption{Numerical results demonstrate the attractive behavior of the obtained fixed point (displayed as a {black point}) for the $\hat{f}(R) = R^{3/2}$ model with $\beta_1>0$. In this context, the chosen parameters are $\beta_1=10^{-4}$ and $\beta_2=-0.04$.}
	\label{fig:numerical-n-1-2-01}
\end{figure}

We begin by investigating the first region, in which $\beta_1>0$ along with
 $B>-\frac{24 \beta_1 (n-2)}{(n-1) n} $. Substituting $n=3/2$ into Eq. \eqref{5.3}, we obtain
\begin{equation}\label{}
	B>16 \beta_1.
\end{equation}
As a result, we derive the following constraint for $\beta_2$,
\begin{equation} \label{stability-constraint-1}
	\beta_2^2>\dfrac{16}{3}\beta_1.
\end{equation}
Hence, if we consider a value for $\beta_1 = 10^{-4}$, we can derive the following constraint for $\beta_2$ such as
\begin{equation}\label{}
	\beta_2  <-\frac{1}{25 \sqrt{3}}.
\end{equation}
It turns out that if we choose $\beta_2 = -1/25 =-0.04 <-1/(25 \sqrt{3})$ then the corresponding value of the coefficient $a_1$ becomes
\begin{equation}\label{}
	a_1 \simeq {-2} <0.
\end{equation}
Apparently, the de Sitter inflationary solution in the $n=3/2$ model is indeed stable as expected. Furthermore, its attractive behavior can also be confirmed by numerical calculations shown in Fig. \ref{fig:numerical-n-1-2-01}. 
\subsubsection{Second region with $\beta_1<0$}
\begin{figure}[hbtp] 
\begin{center}
	\includegraphics[scale=0.45]{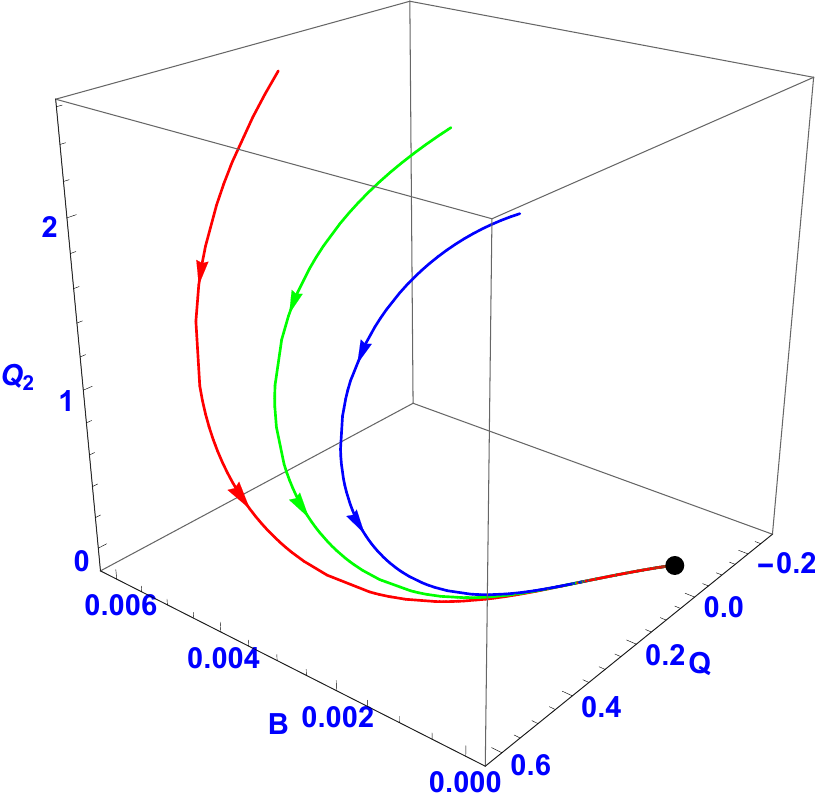}\\
	\caption{Numerical results illustrate the attractive behavior of the obtained fixed point (represented as a {black point}) for the $\hat{f}(R) = R^{3/2}$ model with $\beta_1<0$. Here, the parameters are selected to be $\beta_1=-10^{-4}$ and $\beta_2=-0.004$. }
	\label{fig:numerical-n-1-2-02}
	\end{center}
\end{figure}

In this case, the corresponding de Sitter inflationary solution will always be stable for all $B\neq 0$. To perform numerical calculations, we will take specific values for parameters such as $\beta_1 = -10^{-4}$ and $\beta_2 = -0.004$. Consequently, we obtain
\begin{equation}\label{}
	a_1 \simeq {-0.039} <0.
\end{equation}
Attractor behavior of the de Sitter inflationary solution can be seen in Fig. \ref{fig:numerical-n-1-2-02}.
\subsection{$\hat{f}(R) = R^{1/2}$ model}
As analyzed in the previous section, e.g., see Table \ref{tab:my-table}, the corresponding fixed point will always be stable if $\beta_1>0$. On the contrary, the corresponding fixed point will be stable for $\beta_1<0$ only when the following inequality,
\begin{equation}\label{}
	B>-\frac{24 \beta_1  (n-2)}{(n-1) n},
\end{equation}
is satisfied. As a result, this inequality becomes 
\begin{equation}\label{5.11}
\beta_2>	-\dfrac{1}{3\sqrt{-3\beta_1 }}
\end{equation}
for $n=1/2$.  Now, we are going to consider two regions with $\beta_1>0$ and $\beta_1<0$.
\subsubsection{First region with $\beta_1>0$}
\begin{figure}[hbtp] 
\begin{center}
	\includegraphics[scale= 0.5]{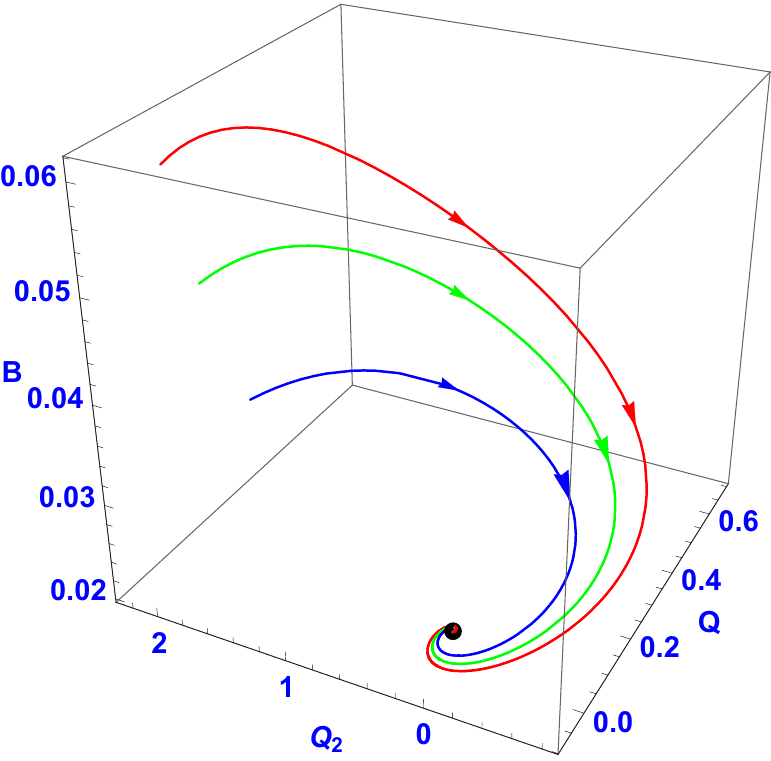}\\
	\caption{The visualization for the stability of the fixed point (displayed as a black point) in the $\hat{f}(R) = R^{1/2}$ model for the case $\beta_1>0$.  The parameters in this scenario are specified as $\beta_1=10^{-4}$ and $\beta_2=-15$. As one can see, with different initial conditions, over time the curves converge to the black fixed point.}
	\label{fig:numerical-n-0-1-01}
	\end{center}
\end{figure}

In this region, the fixed point is always stable. As an illustration, we take the specific values of parameters such as $\beta_1 = 10^{-4}$ and $\beta_2 = -15$. As a result, the corresponding value of the coefficient $a_1$ is given by
\begin{equation}\label{}
	a_1 \simeq {-7.47}<0.
\end{equation}
Hence, it is evident that the fixed point is indeed stable. Interestingly, numerical result shown in Fig. \ref{fig:numerical-n-0-1-01} does confirm this result.
\subsubsection{Second region with $\beta_1<0$}
\begin{figure}[hbtp] 
\begin{center}
	\includegraphics[scale= 0.5]{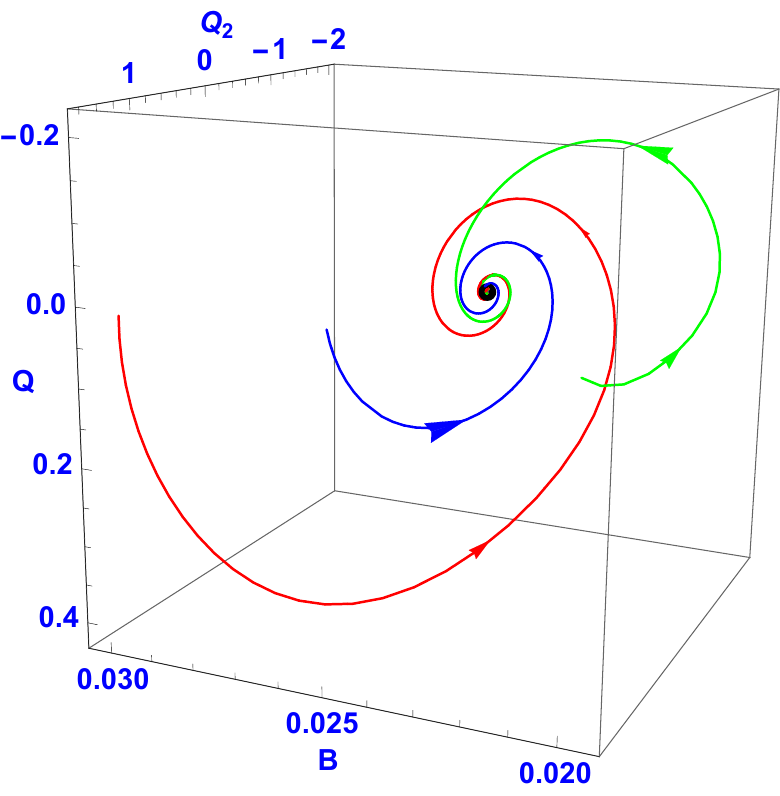}\\
	\caption{Numerical results confirm the attractive behavior of the obtained fixed point, represented as a black point, in the $\hat{f}(R) = R^{1/2}$ model for the case $\beta_1<0$. Here, the parameters are specified as $\beta_1=-10^{-4}$ and $\beta_2=-15$.}
	\label{fig:numerical-n-0-1-02}
	\end{center}
\end{figure}

In this region,  a stable fixed point will be archived once the constraint \eqref{5.11} is fulfilled. For instance, if we take $\beta_1 = -10^{-4}$, then we can choose $\beta_2 = -15 > -{1}/(3\sqrt{-3\beta_1 }) \simeq -19.25$ in order to satisfy the inequality \eqref{5.11}. With these parameter values, we can determine the following value of $a_1$ to be
\begin{equation}\label{}
	a_1 \simeq {-30.57} <0.
\end{equation}
Hence, the fixed point is indeed stable. Furthermore, its numerical support can be seen in Fig. \ref{fig:numerical-n-0-1-02}.
\subsection{$\hat{f}(R) = R^{-1}$ model}
\begin{figure}[hbtp] 
\begin{center}
	\includegraphics[scale= 0.5]{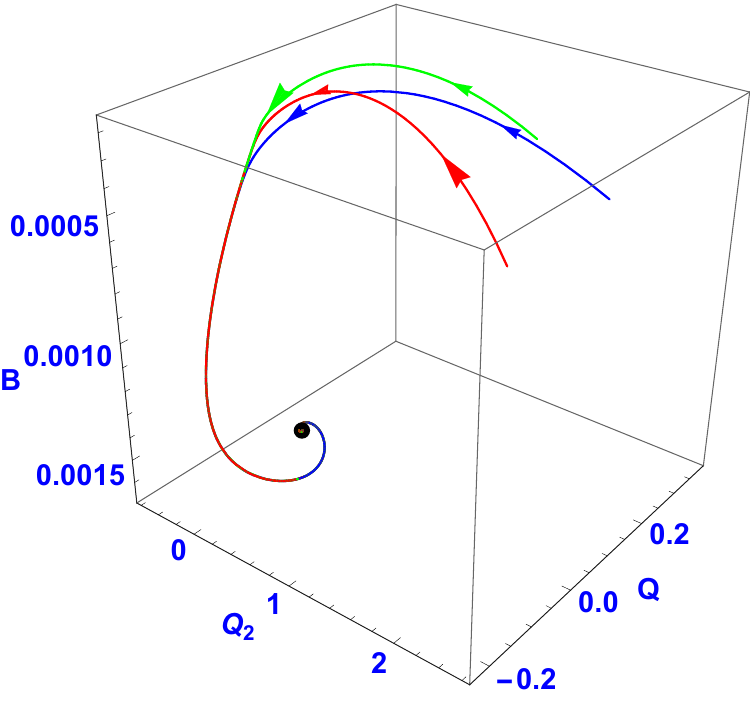}\\
	\caption{Numerical results confirm the attractive behavior of the obtained fixed point (displayed as a black point) for the $\hat{f}(R) = R^{-1}$ model for the case $\beta_1>0$.  In this scenario,  the parameters are setting as $\beta_1=10^{-4}$ and $\beta_2=-2\times 10^7$.}
	\label{fig:numerical-n-0}
	\end{center}
\end{figure}

Finally, we examine a concrete model with $n=-1$ in the range of $n<0$. As observed in Table \ref{tab:my-table}, the  stability of the fixed point is achieved solely in the region with $\beta_1>0$ along with the following constraint  
\begin{equation}\label{}
	B<-\frac{24 \beta_1  (n-2)}{(n-1) n}.
\end{equation}
As a result, the corresponding condition for the parameter $\beta_2$ turns out to be
\begin{equation}\label{}
	\beta_2<-\frac{1}{27 \beta_1 ^2}.
\end{equation}
By setting $\beta_1 = 10^{-4}$, this condition becomes
\begin{equation}\label{}
	\beta_2 <-3.7037\times 10^{6}.
\end{equation}
Therefore, we will choose a specific value as $\beta_2 = -2\times 10^7$ to fulfill this condition, and subsequently, we can compute the corresponding value of $a_1$ to be
\begin{equation}\label{key}
	a_1\simeq {-9.06}<0.
\end{equation}
Hence, we can conclude that the corresponding fixed point is indeed stable. This conclusion is further validated by numerical results shown in Fig. \ref{fig:numerical-n-0}.
\section{Stability analysis extended: Effective potential method} \label{sec7}
In this section, we will extend our stability analysis by using an effective potential method, which has been extensively used by other people to investigate the stability of de Sitter solutions, e.g., see Refs. \cite{Ivanov:2011np,Skugoreva:2014gka,Elizalde:2014xva,Pozdeeva:2019agu,Vernov:2021hxo}. Basically, this method can be applied by considering the following action \cite{Sotiriou:2008rp,DeFelice:2010aj}
\begin{equation} \label{action-extened}
S= \frac{M_p^2}{2} \int d^4x \sqrt{-g} \left [ R + f'(\phi)(R-\phi)+ f(\phi)\right],
\end{equation} 
where $f(\phi)$ is a double differentiable function of $\phi$ and $f'(\phi)\equiv df/d\phi$. Varying this action with respect to $\phi$, we obtain
\begin{equation}
f''(\phi) (R-\phi)=0.
\end{equation}
Assuming that $f''(\phi) \neq 0$, then this equation implies 
\begin{equation}
\phi =R.
\end{equation}
Inserting this solution into Eq. \eqref{action-extened}, we obtain the corresponding action of $R+f(R)$ gravity,
\begin{equation} \label{action-extened-fR}
S= \frac{M_p^2}{2} \int d^4x \sqrt{-g} \left [ R +f(R)\right].
\end{equation} 
Furthermore, if we define
\begin{equation}
\varphi = f'(\phi),
\end{equation}
then the action \eqref{action-extened} becomes that of the so-called Brans-Dicke theory with a potential of scalaron $\varphi$ \cite{Sotiriou:2008rp,DeFelice:2010aj}
\begin{equation} \label{BD}
S= \int d^4x \sqrt{-g} \left [\frac{M_p^2}{2} \varphi R - V(\varphi)\right],
\end{equation} 
where $V(\varphi)$ is the potential of scalaron defined as
\begin{equation}
V(\varphi) = \frac{M_p^2}{2} \left[ \phi (\varphi) \left(\varphi -1 \right) -f\left(\phi(\varphi) \right)\right].
\end{equation}
In our present paper, we have been interested in power-law extensions of the Starobinsky model with the corresponding function $f$ given by
\begin{equation}
f(R) = \beta_1 R^2 + \beta_2 R^n,
\end{equation}
with $n \neq 0, ~1$, and $2$. This implies that
\begin{equation}\label{7.8}
f(\phi) = \beta_1 \phi^2 + \beta_2 \phi ^n
\end{equation}
along with
\begin{equation} \label{relation-phi-varphi}
\varphi = f'(\phi) = 2\beta_1 \phi + n\beta_2 \phi^{n-1}.
\end{equation}
In principle, solving Eq. \eqref{relation-phi-varphi} will yield solutions of $\phi(\varphi)$ along with the corresponding form of potential $V(\varphi)$. Generically, Eq. \eqref{relation-phi-varphi} will be very complicated to be solved analytically when $n \gg 1$. However, our current consideration has been limited to some small values of $n$  such as $n=3, ~3/2, ~1/2$, and $-1$, such that Eq.  \eqref{relation-phi-varphi} can be solved analytically. 

Now, we will investigate whether the action \eqref{BD} admits stable de Sitter inflationary solutions. First, the corresponding field equations of this model in the FLRW metric \eqref{2} with $N(t)=1$ are given by \cite{Ivanov:2011np,Skugoreva:2014gka,Elizalde:2014xva,Pozdeeva:2019agu,Vernov:2021hxo}
\begin{align}
\label{Frid-1}
3M_p^2 \dot\alpha \left(\varphi \dot\alpha +\dot\varphi \right) -V =&~0,\\
\label{Frid-2}
M_p^2 \left[ \varphi \left( 3\dot\alpha^2 +2\ddot\alpha \right) +\ddot\varphi +2\dot\varphi \dot\alpha \right] -V=&~0,\\
\label{Frid-3}
3M_p^2 \left(\ddot\alpha+ 2\dot\alpha^2  \right)-V'=&~0.
\end{align}
Similar to the previous section, we are looking for the de Sitter solutions with 
\begin{equation}
\alpha=\zeta t,\quad \dot\alpha =\zeta,\quad \ddot\alpha =0.
\end{equation}
As a result, 
Eqs. \eqref{Frid-1} and \eqref{Frid-2} imply that
\begin{equation}
\ddot\varphi - \zeta \dot\varphi =0,
\end{equation}
which admits a compatible solution $\dot\varphi = 0$, or equivalently, $\varphi =\varphi_{\rm dS} ={\rm constant}$. Note that another time-dependent solution, $\varphi(t) \sim e^{\zeta t}/\zeta$ has been ignored since it would lead to a
varying effective gravitational constant proportional to $1/\varphi(t)$ and having properties different from the standard de Sitter solutions as stated in Ref. \cite{Pozdeeva:2019agu}.  Furthermore, $\dot\varphi=0$ implies that $\dot\phi =0$ as expected for the de Sitter solutions. This solution is nothing but that we have considered in the previous section for seeking the de Sitter solutions. Interestingly, $\zeta$ can be determined, according to Eq. \eqref{Frid-1}, such  as
\begin{equation} \label{solution-BD}
\zeta = \sqrt{\frac{V}{3M_p^2\varphi_{\rm dS}}}.
\end{equation}
Hence, $\zeta \gg 1$ will require $|V| \gg |3M_p^2 \varphi_{\rm dS}|$, while $\zeta \sim 1$ will correspond to $|V| \sim |3M_p^2 \varphi_{\rm dS}|$.  Additionally, $\varphi_{\rm dS}$ must be positive definite for the positive $V$. Otherwise, it should be negative definite.  To investigate the stability of the obtained de Sitter inflationary solutions, we transform the field equations \eqref{Frid-1}, \eqref{Frid-2}, and \eqref{Frid-3} into the corresponding dynamical system by introducing dynamical variables such as \cite{Ivanov:2011np,Skugoreva:2014gka,Elizalde:2014xva,Pozdeeva:2019agu,Vernov:2021hxo}
\begin{equation}
x = \dot\alpha, \quad y = \varphi,\quad z=\dot\varphi.
\end{equation}
As a result, the corresponding set of autonomous equations is defined to be
\begin{align}
	\frac{d x}{d\alpha}& = \frac{V'}{3 M_p^2 x}-2 x,\\
	\frac{d y}{d\alpha}& =\frac{z}{x},\\
	\frac{d z}{d\alpha}& = -\frac{2 y V'}{3 M_p^2 x}+\frac{V}{M_p^2 x}+x y-2 z.
\end{align}
It is clear that a non-trivial fixed point of this dynamical system is given by 
\begin{align}\label{fixed-point-BD-1}
	x^2&=\frac{V'}{6M_p^2}, \\
	\label{fixed-point-BD-2}
	z&=0, \\
	\label{fixed-point-BD-3}
	y&=\frac{2 V}{V'},
\end{align}
which is derived from a set of equations, $dx/d\alpha =dy/d\alpha =dz/d\alpha=0$. It is straightforward to verify that this fixed point is indeed equivalent to the de Sitter solution shown in Eq. \eqref{solution-BD}.

Now, we are interested in the stability of this fixed point during the inflationary phase. As a result, perturbed equations from the autonomous equations are given by
\begin{align}
	\frac{d\delta x}{d\alpha}& =  \left(-\frac{V'}{3 M_p^2 x^2}-2\right)\delta x + \frac{ V''}{3 M_p^2 x}\delta y,\\
	\frac{d\delta y}{d\alpha}& = -\frac{z}{x^2}\delta x +\frac{1}{x} \delta z ,\\
	\frac{d\delta z}{d\alpha}& =  \left(\frac{2 y V'}{3 M_p^2 x^2}-\frac{V}{M_p^2 x^2}+y\right)\delta x +  \left(-\frac{2 y V''}{3 M_p^2 x}+\frac{V'}{3 M_p^2 x}+x\right)\delta y -2\delta z.
\end{align}
By taking exponential perturbations such as
\begin{equation}
\delta x =A_1 e^{\kappa \alpha},\quad \delta y =A_2 e^{\kappa \alpha},\quad \delta z =A_3 e^{\kappa \alpha},
\end{equation}
we can rewrite the perturbed equations as a matrix equation,
	\begin{align}\label{}
		\mathcal{\hat M}  \begin{pmatrix}
			A_1\\
			A_2\\
			A_{3}
		\end{pmatrix}&\equiv	
		\begin{bmatrix}
			-\frac{V'}{3 M_p^2 x^2} -2  -\kappa & \frac{V''}{3 M_p^2 x} & 0 \\
			-\frac{z}{x^2} & -\kappa & \frac{1}{x}  \\
			\frac{2 y V'}{3 M_p^2 x^2}-\frac{V}{M_p^2 x^2}+y & -\frac{2 y V''}{3 M_p^2 x}+\frac{V'}{3 M_p^2 x}+x &-2-\kappa\\
		\end{bmatrix}
		\begin{pmatrix}
			A_1\\
			A_2\\
			A_{3}
		\end{pmatrix}=0.
	\end{align}
This homogeneous equation admits non-trivial solutions of $A_i$ if and only if
\begin{equation}
\det \mathcal{\hat M} =0,
\end{equation}
which can be expanded to be a polynomial equation of $\kappa$ such as
\begin{equation} \label{equation of kappa}
	a_3\kappa^3+a_2 \kappa^2+a_1 \kappa +a_0=0,
\end{equation}
with
\begin{align}
	a_3 &=-1,\\
	a_2 &=-6,\\
	a_1& =-5-\frac{2 y V''}{3 M_p^2 x^2},\\
	a_0&= 12-\frac{2 y V''}{M_p^2 x^2}.
\end{align}
 Here, all coefficients are taken at the de Sitter fixed point shown in Eqs. \eqref{fixed-point-BD-1}, \eqref{fixed-point-BD-2}, and \eqref{fixed-point-BD-3}. 
Following Refs. \cite{Ivanov:2011np,Skugoreva:2014gka,Elizalde:2014xva,Pozdeeva:2019agu,Vernov:2021hxo}, we define an effective potential $V_{\rm eff}$ such as
\begin{equation}\label{7.36}
V_{\rm eff} =-\frac{\varphi^2}{V}.
\end{equation}
It is clear that $V_{\rm eff} V<0$ for a real $\varphi$. It is noted that the effective potential \eqref{7.36} is not a unique function appropriate for  investigations of the stability of the de Sitter solutions \cite{Elizalde:2014xva,Pozdeeva:2019agu,Vernov:2021hxo}. For instance, one may consider an alternative function such as $ \tilde{V}_{\rm eff} =-1/V_{\rm eff} = V/\varphi^2$  \cite{Elizalde:2014xva}.

From Eq. \eqref{fixed-point-BD-3}, we have
\begin{equation}
	\frac{V}{V'} =\frac{\varphi}{2},
\end{equation}
which is nothing but a non-trivial solution of $V'_{\rm eff}=0$. This implies that the de Sitter fixed point corresponds to $V'_{\rm eff}=0$. Furthermore,  it is easy to obtain from Eq. \eqref{7.36} that
\begin{equation}
	V''=-\frac{4 V'}{y}+\frac{2 V'^2}{V}+\frac{V^2 V_{\rm eff}''}{y^2}+\frac{2 V}{y^2}.
\end{equation}
For the de Sitter fixed point, we have
\begin{equation}
	V'' = 9 M_p^4 x^4 V''_{\rm eff}+\frac{6 M_p^2 x^2}{y},
\end{equation}
by which $a_1$ and $a_0$ reduce to
\begin{align} \label{defintion of a1}
a_1 &= -9 - 6 M_p^2 x^2 y V''_{\rm eff},\\
a_0 &=-18 M_p^2 x^2 y V''_{\rm eff},
\end{align}
respectively. It becomes clear that Eq. \eqref{equation of kappa} will always admit at least one positive root $\kappa>0$ if $a_0>0$, or equivalently $V''_{\rm eff} <0$ provided that $y>0$,  since $a_3=-1<0$. Therefore, in order to avoid any unstable de Sitter fixed point, it must require that 
\begin{equation}
V''_{\rm eff} >0.
\end{equation}
Interestingly, the positivity of $V''_{\rm eff} $ will result the negativity of $a_1$ according to Eq. \eqref{defintion of a1}, which will ensure the existence of all negative roots $\kappa<0$ of Eq. \eqref{equation of kappa}. In other words, the condition $V''_{\rm eff} >0$ will correspond to the stable de Sitter fixed point. By noting that the de Sitter fixed point is a solution of $V'_{\rm eff} =0$ as shown above, we come to an important conclusion that the stable de Sitter fixed point is exactly  the minima of the effective potential $V_{\rm eff}$. In sort, if we are able to confirm the existence of such minima, we will certainly derive the corresponding stable de Sitter inflationary solutions for our considered models. Our analysis is indeed consistent with the previous ones by other people in Refs. \cite{Ivanov:2011np,Skugoreva:2014gka,Elizalde:2014xva,Pozdeeva:2019agu,Vernov:2021hxo}.

We  are now considering a specific case of our current model with $n = 3$, i.e., $\hat f(R)=R^3$ as a demonstration. First, we find a suitable solution of $\phi(\varphi)$ by solving Eq. \eqref{relation-phi-varphi},
\begin{equation}
	\phi (\varphi) = \frac{\sqrt{3 \beta_2 \varphi +\beta_1^2}-\beta_1}{3 \beta_2}.
\end{equation}
Another solution of Eq. \eqref{relation-phi-varphi} has been neglected since it implies $\varphi_{\rm dS} >0$ and $V(\varphi_{\rm dS}) <0$, which really violates the condition for the existence of real $\zeta$, whose definition has been shown in Eq. \eqref{solution-BD}. This can be observed by plotting the corresponding $V_{\rm eff}$ to see the sign of $V$ due to the fact that $V_{\rm eff} V<0$ as shown above.

Consequently, the following expression for $f(\phi(\varphi))$ can be obtained, according to Eq. \eqref{7.8}, such as
\begin{equation}
	f(\phi (\varphi)) = \frac{ \left(\sqrt{3 \beta _2 \varphi +\beta _1^2}-\beta _1\right)^2 \left(\sqrt{3 \beta _2 \varphi +\beta _1^2}+2 \beta _1\right)}{27 \beta _2^2}.
\end{equation}
As a result, the corresponding potential $V(\varphi)$ reads
\begin{equation}
	V(\varphi) =- \frac{M_p^2 \left(\sqrt{3 \beta _2 \varphi +\beta _1^2}-\beta _1\right) \left[\beta _1 \left(\sqrt{3 \beta _2 \varphi +\beta _1^2}-\beta _1\right)+\beta _2 (9-6 \varphi )\right]}{54 \beta _2^2},
\end{equation}
which leads to the final form of the effective potential $V_{\rm eff}(\varphi)$,\begin{equation}
	V_{\rm eff}(\varphi) = \frac{54 \beta _2^2 \varphi ^2}{M_p^2 \left(\sqrt{3 \beta _2 \varphi +\beta _1^2}-\beta _1\right) \left[\beta _1 \left(\sqrt{3 \beta _2 \varphi +\beta _1^2}-\beta _1\right)+\beta _2 (9-6 \varphi )\right]}.
\end{equation}
Then, the first derivative of the effective potential, i.e., $V'_{\rm eff}(\varphi)$, is given by
\begin{align}
	V'_{\rm eff}(\varphi) =&~ \frac{27 \beta _2^2 \varphi  }{M_p^2 \sqrt{3 \beta _2 \varphi +\beta _1^2}  \left(\sqrt{3 \beta _2 \varphi +\beta _1^2}-\beta _1\right)^2 \left[\beta _1 \left(\sqrt{3 \beta _2 \varphi +\beta _1^2}-\beta _1\right)+\beta _2 (9-6 \varphi )\right]^2}\nonumber\\
	&\times\left[8 \beta _1^3 \sqrt{3 \beta _2 \varphi +\beta _1^2}+6 \beta _2 \beta _1^2 (6-5 \varphi )+18 \beta _2 \beta _1 (\varphi -2) \sqrt{3 \beta _2 \varphi +\beta _1^2}\right.\nonumber\\
	&~\quad\left.+9 \beta _2^2 \varphi  (9-2 \varphi )-8 \beta _1^4\right].
\end{align}
Consequently, it is easy to define $V''_{\rm eff}(\varphi)$ to be
\begin{align}
	V''_{\rm eff}(\varphi) =~&\frac{27 \beta _2^2}{2 M_p^2 \left(3 \beta _2 \varphi +\beta _1^2\right){}^{3/2} \left(\sqrt{3 \beta _2 \varphi +\beta _1^2}-\beta _1\right)^3 \left[\beta _1 \left(\sqrt{3 \beta _2 \varphi +\beta _1^2}-\beta _1\right)+\beta _2 (9-6 \varphi )\right]^3 }\nonumber\\
	&\times\left\{243 \beta _2^4 \beta _1 \varphi ^2 [\varphi  (6 \varphi -19)-27]-81 \beta _2^4 \varphi ^2 [4 (\varphi -9) \varphi -27] \sqrt{3 \beta _2 \varphi +\beta _1^2}\right.\nonumber\\
	&\left.+64 \beta _1^8 \sqrt{3 \beta _2 \varphi +\beta _1^2}-288 \beta _2 \beta _1^7 (\varphi -2)+192 \beta _2 \beta _1^6 (\varphi -3) \sqrt{3 \beta _2 \varphi +\beta _1^2}\right.\nonumber\\
	&\left.+108 \beta _2^2 \beta _1^5 [\varphi  (\varphi +24)-12]-108 \beta _2^2 \beta _1^4 (\varphi +6) (3 \varphi -2) \sqrt{3 \beta _2 \varphi +\beta _1^2}\right.\nonumber\\
	&\left.+54 \beta _2^3 \beta _1^3 \varphi  [\varphi  (31 \varphi +18)-108]+108 \beta _2^3 \beta _1^2 \varphi  [\varphi  (9-10 \varphi )+36] \sqrt{3 \beta _2 \varphi +\beta _1^2}-64 \beta _1^9\right\}.
\end{align}

\begin{figure}[hbtp] 
\begin{center}
	\includegraphics[scale= 0.7]{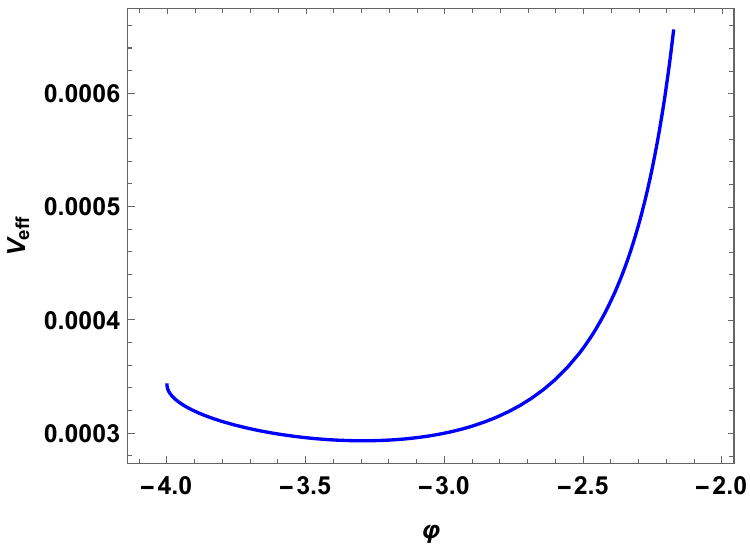}\\
	\caption{Effective potential $V_{\rm eff}$ as a function of $\varphi$ for $n=3$ and $M_p=1$. }
	\label{minimum}
	\end{center}
\end{figure}

Note that the condition $V'_{\rm eff}(\varphi_{\rm dS}) = 0$ yields the de Sitter fixed point $\varphi_{\rm dS}$, whose analytical form is highly complex due to the nonlinearity of this equation.  However, for fixed values of $\beta_1$ and $\beta_2$, this equation is not difficult to solve. In particular, with $\beta_1 = -10^{-4}$ and $\beta_2 = \beta_1^2/12$ as chosen above for the inflationary solution, we are able to obtain an approximate value of $\varphi_{\rm dS}$ such as  $\varphi_{\rm dS} \simeq -3.29$ by solving equation $V'_{\rm eff}(\varphi_{\rm dS}) = 0$. Thanks to this value, the corresponding value of $V''_{\rm eff}(\varphi_{\rm dS})$ turns out to be
\begin{equation}
	V''_{\rm eff}(\varphi_{\rm dS})\simeq \dfrac{1.35 \times 10^{-4}}{M_p^2}>0.
\end{equation}
As stated above, this positivity of $V''_{\rm eff}(\varphi_{\rm dS})$ really confirms the stability of the de Sitter fixed point during the inflationary phase. It also confirms that this stable de Sitter fixed point is a minimum of $V_{\rm eff}$ (see Fig. \ref{minimum} for details). More importantly, this result is indeed consistent with the another stability analysis presented in the subsection \ref{subsection-R3} for $\hat f(R)=R^3$. Similar conclusions for the other cases of $n$ such as $n=3/2, ~1/2$, and $-1$ could be obtained accordingly. 
To demonstrate and visualize this claim, we plot the corresponding effective potential $V_{\rm eff}$ for these cases. See Fig. \ref{fig:3/2, 1/2 and -1} for details. It is evident from these plots that all considered cases of $n$ always yield desired minima of the effective potential $V_{\rm eff}$, indicating that the corresponding de Sitter solutions are indeed stable during the inflationary phase. This result once again confirms the validity of the stability analysis done in the previous section. To end this section, it is worth noting that the small values of $|V_{\rm eff}(\varphi_{\rm dS})|$ compared to that of $|\varphi_{\rm dS}|$ with $M_p=1$ are definitely consistent with the inflationary solutions with $\zeta \gg 1$. 
\begin{figure}[htp!]
		\begin{subfigure}[b]{0.3\textwidth}
	\includegraphics[scale=0.4]{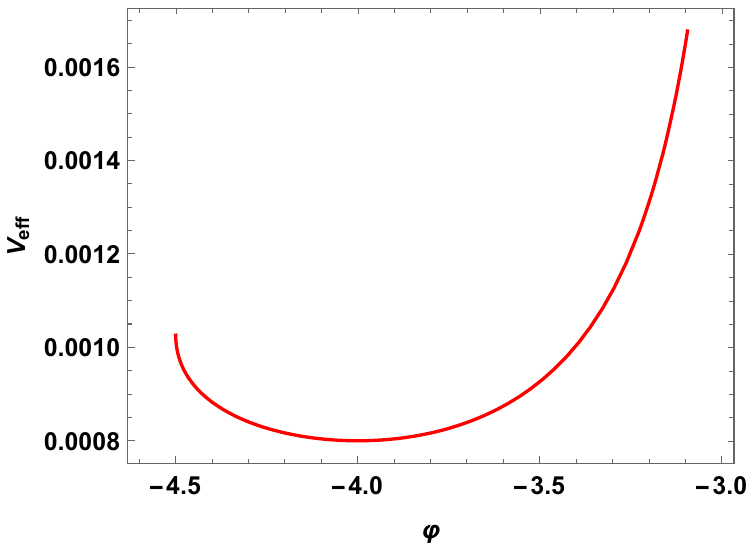}
	\end{subfigure}
		\qquad
	\begin{subfigure}[b]{0.3\textwidth}
		\includegraphics[scale=0.4]{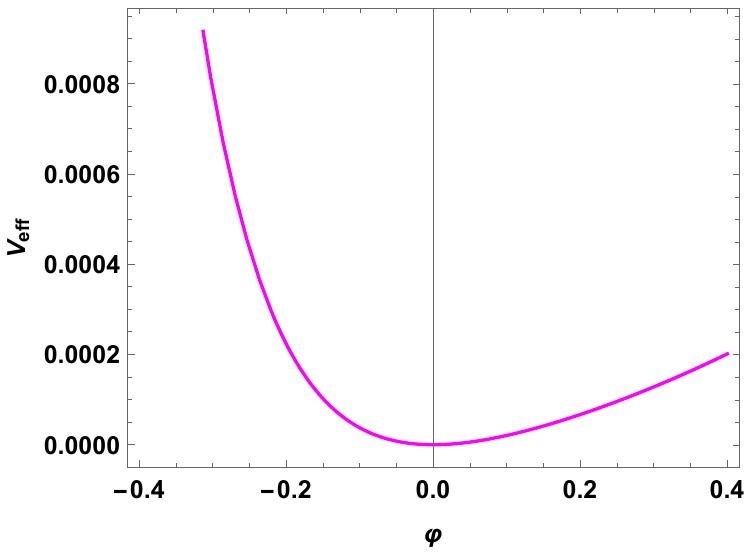}
	\end{subfigure}
	\quad 
	\begin{subfigure}[b]{0.3\textwidth}
		\includegraphics[scale=0.4]{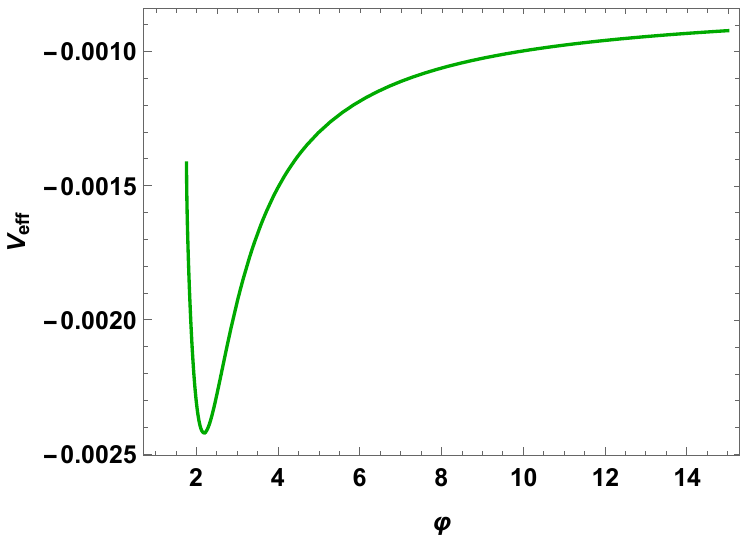}
	\end{subfigure}
	\caption{(From left to right) Effective potential $V_{\rm eff}$ as a function of $\varphi$ for $n=3/2,~ 1/2,$ and $-1$, respectively. $M_p=1$ is also chosen for simplicity. The values of parameters $\beta_1$ and $\beta_2$ for each case of $n$ have been employed from the previous sections.}
	\label{fig:3/2, 1/2 and -1}
\end{figure}
\section{Conclusions}\label{final}
We have investigated whether stable de Sitter inflationary solutions appear within power-law extensions of the Starobinsky model. In particular, we will address general constraints for the existence along with the stability of de Sitter inflationary solutions in a general case involving not only the Starobinsky $R^2$ term but also an additional $R^n$ one. For convenience, we have summarized in the Table \ref{tab:my-table} all possibilities to obtain stable de Sitter inflationary solutions in the framework of power-law extensions of the Starobinsky model. According to this table, we could figure out whether the inflationary phase is compatible with the power-law extensions of the Starobinsky model, following the discussions of recent papers \cite{Pozdeeva:2019agu,Vernov:2021hxo}. In particular, the extensions having saddle fixed points, which are equivalent to unstable de Sitter inflationary solutions, would be more suitable for the inflationary phase of the early universe. On the other hand, the extensions admitting  attractor fixed points, which are equivalent to stable de Sitter inflationary solutions, would be more suitable for the accelerated expansion of the late time universe due to the so-called graceful exit problem. Specific values of $n$ such as $n=3, ~3/2, ~1/2$, and $-1$ have been chosen for transparent demonstrations. It is worth noting that the authors of Ref.  \cite{Ivanov:2021chn} have arrived at an important conclusion that unlike the $R^3$ and $R^4$  terms, the term $R^{3/2}$ has a significant impact on the value of tensor-to-scalar ratio. Therefore, CMB imprints of the power-law extensions of the Starobinsky model will be our future studies. We hope that our present study could provide useful information for cosmological implications of non-trivial extensions of the Starobinsky model.

\begin{acknowledgments}
We would like to thank a referee very much for useful comments and suggestions.
\end{acknowledgments}

\end{document}